\documentclass[useAMS,usenatbib]{mn2e}
\usepackage[colorlinks,citecolor=blue]{hyperref}
\usepackage{amsmath}
\usepackage{graphicx}
\usepackage{epstopdf}
\usepackage{amssymb}
\usepackage{extarrows}
\usepackage{bm}
\usepackage{float}
\usepackage{color}
\newcommand{\ba}{\begin{eqnarray}}
\newcommand{\ea}{\end{eqnarray}}
\newcommand{\be}{\begin{equation}}
\newcommand{\ee}{\end{equation}}
\newcommand{\gr}{\mathrm{GR}}

\newcommand{\m}{\mathrm{max}}
\newcommand{\mi}{\mathrm{min}}

\newcommand{\au}{\mathrm{AU}}

\newcommand{\IN}{\mathrm{in}}
\newcommand{\MID}{\mathrm{mid}}
\newcommand{\OUT}{\mathrm{out}}

\newcommand{\lk}{\mathrm{LK}}
\newcommand{\merger}{\mathrm{merger}}
\newcommand{\eff}{\mathrm{eff}}

\newcommand{\li}{\mathrm{lim}}

\def\e1{e_1^2}

\title[Hierarchical Mergers in Multiple Systems]
{Hierarchical Black-Hole Mergers in Multiple Systems: Constrain the Formation of GW190412, GW190814 and GW190521-like events}

\author[Liu, \& Lai]
{Bin Liu$^{1}$, Dong Lai$^{1,2}$\\
$^{1}$ Cornell Center for Astrophysics and Planetary Science, Department of Astronomy, Cornell University, Ithaca, NY 14853, USA\\
$^{2}$ Tsung-Dao Lee Institute, Shanghai Jiao Tong University, Shanghai 200240, China\\
}

\begin{document}


\pagerange{\pageref{firstpage}--\pageref{lastpage}} \pubyear{2020}

\maketitle

\label{firstpage}

\begin{abstract}
{The merging black-hole (BH) binaries GW190412, GW190814 and
    GW190521 from the third LIGO/VIRGO observing run exhibit some
    extraordinary properties, including highly asymmetric masses,
    significant spin, and component mass in the ``mass gap''.  These
    features can be explained if one or both components of the binary
    are the remnants of previous mergers. In this paper, we explore
    hierarchical mergers in multiple stellar systems, taking into
    account the natal kick and mass loss due to the supernova explosion
    (SN) on each component, as well as the merger kick received by the
    merger remnant. The binaries that have survived the SNe and kicks
    generally have too wide orbital separations to merge by
    themselves, but can merge with the aid of an external companion
    that gives rise to Lidov-Kozai oscillations.
    The BH binaries that consist of second-generation BHs
    can also be assembled in dense star clusters through binary interactions.
    We characterize the parameter space of these BH binaries by merger
    fractions in an analytical approach. Combining the distributions
    of the survived binaries, we further constrain the parameters of
    the external companion, using the analytically formulated tertiary
    perturbation strength. We find that to produce the three
    LIGO/VIRGO O3 events, the external companions must be at least a
    few hundreds $M_\odot$, and fall in the intermediate-mass BH and
    supermassive BH range. We suggest that GW190412, GW190814 and
    GW190521 could all be produced via hierarchical mergers in multiples,
    likely in a nuclear star cluster, with the final merger induced by a
    massive BH.
}
\end{abstract}


\section{Introduction}
\label{sec 1}

The detections of gravitational waves from merging
binary black holes (BHs) \citep[e.g.,][]{LIGO 2019a,LIGO 2019b} have motivated many
recent studies on their formation channels, including the
traditional isolated binary evolution
\citep[e.g.,][]{Lipunov 1997,Lipunov 2017,Podsiadlowski 2003,
Belczynski 2010,Belczynski 2016,Dominik 2012, Dominik 2013,Dominik 2015},
the chemically homogeneous evolution \citep[e.g.,][]{Mandel and de Mink 2016,Marchant 2016},
the gas-assisted mergers in AGN disks \citep[e.g.,][]{Bartos},
and various flavors of dynamical channels that involve either strong gravitational
scatterings in dense clusters
\citep[e.g.,][]{Portegies 2000,O'Leary 2006,Miller 2009,Banerjee 2010,Downing 2010,Ziosi 2014,Rodriguez 2015,Samsing 2018}
or Lidov-Kozai induced mergers in isolated triple and quadruple systems
\citep[e.g.,][]{Blaes 2002,Miller 2002,Wen 2003,Antonini 2012,Antonini 2017,Silsbee and Tremaine 2017,
Hoang 2017,Liu APJ,Liu Quad,Liu SMBH 2,Xianyu 2018,Liu APJ 2,Liu SMBH,Fragione 2019a,Fragione 2019b}.

The BH mergers detected in the first and second runs (O1 and O2) of LIGO/VIRGO typically feature binaries
with comparable masses (i.e., $m_2/m_1\sim0.6-0.9$) and mass-weighted effective spin parameter $\chi_\eff$
consistent with $\chi_\eff\sim 0$ \citep[but see][]{Zackay 2019,Venumadhav 2020}.
However, the recent detections in the third observing run (O3) of LIGO/VIRGO
reveal the existence of different types of black hole binaries (BHBs).
In GW190412 \citep[][]{GW190412}, the BHB has component masses $29.7^{+5.0}_{-5.3}M_\odot$ and $8.4^{+1.7}_{-1.0}M_\odot$.
The primary (more massive) BH is inferred to rotate rapidly, with the dimensionless spin (Kerr) parameter $\chi_1=0.43^{+0.16}_{-0.26}$.
The effective spin parameter of the BHB is constrained to be $\chi_\eff=0.25^{+0.09}_{-0.11}$, indicating a non-neglgible spin-orbit misalignment
angle.
In the second event GW190814 \citep[][]{GW190814}, the merging binary involves a $23.2^{+1.1}_{-1.0}M_\odot$ BH and a compact
object with mass $2.6^{+0.08}_{-0.09}M_\odot$ in the so-called low mass gap
of ($2M_\odot-5M_\odot$) \citep[e.g.,][]{Bailyn 1998,Feryal 2010,Farr 2011};
the secondary could be a heavy neutron star (NS) or a light BH.
In this source, the primary spin is tightly constrained to a small value ($\chi_1 \lesssim0.07$),
while the secondary spin is unconstrained.
In GW190521\citep[][]{GW190521}, the two BHs have masses of $85^{+21}_{-14}M_\odot$ and $66^{+17}_{-18}M_\odot$, both of which may fall in
the high mass gap produced by the (pulsational) pair-instability supernova processes ($65M_\odot-120M_\odot$)
\citep[e.g.,][]{Barkat 1967,Woosley 2017}. The
analyses of GW190521 indicate that the two BHs are fast-rotating with
$\chi_1=0.69^{+0.27}_{-0.62}$ and $\chi_2=0.73^{+0.24}_{-0.64}$,
while the binary has an effective spin $\chi_\eff=0.08^{+0.27}_{-0.36}$
and ``perpendicular" spin $\chi_\mathrm{p}=0.68^{+0.25}_{-0.37}$, again suggesting significant spin-orbit misalignments.

The astrophysical origin of these three LIGO events are still under debate.
The mergers of BHBs with extreme mass ratios (as in GW190412 and GW190814) or component masses in the mass gap (as in GW190521)
are expected to be rather uncommon, especially if the merging binary contains two
first generation (1G) BHs \citep[e.g.,][]{Gerosa 2020}.
Indeed, recent studies suggest that GW190412- and/or GW190814-like events
should be rare in many formation models, including isolated binaries with
tidally spun-up secondaries \citep[e.g.,][]{Olejak 2020,Mandel 2020}, dynamical assembly
in dense star clusters \citep[e.g.,][]{Di Carlo 2020,Rodriguez 2020,Zevin 2020,Samsing 2020},
gas-assisted formation in AGN disks \citep[e.g.,][]{Yang 2020},
as well as the triple/quadruple stars \citep[e.g.,][]{Hamers 2020,Lu 2021}.

\begin{figure}
\centering
\begin{tabular}{cccc}
\includegraphics[width=9cm]{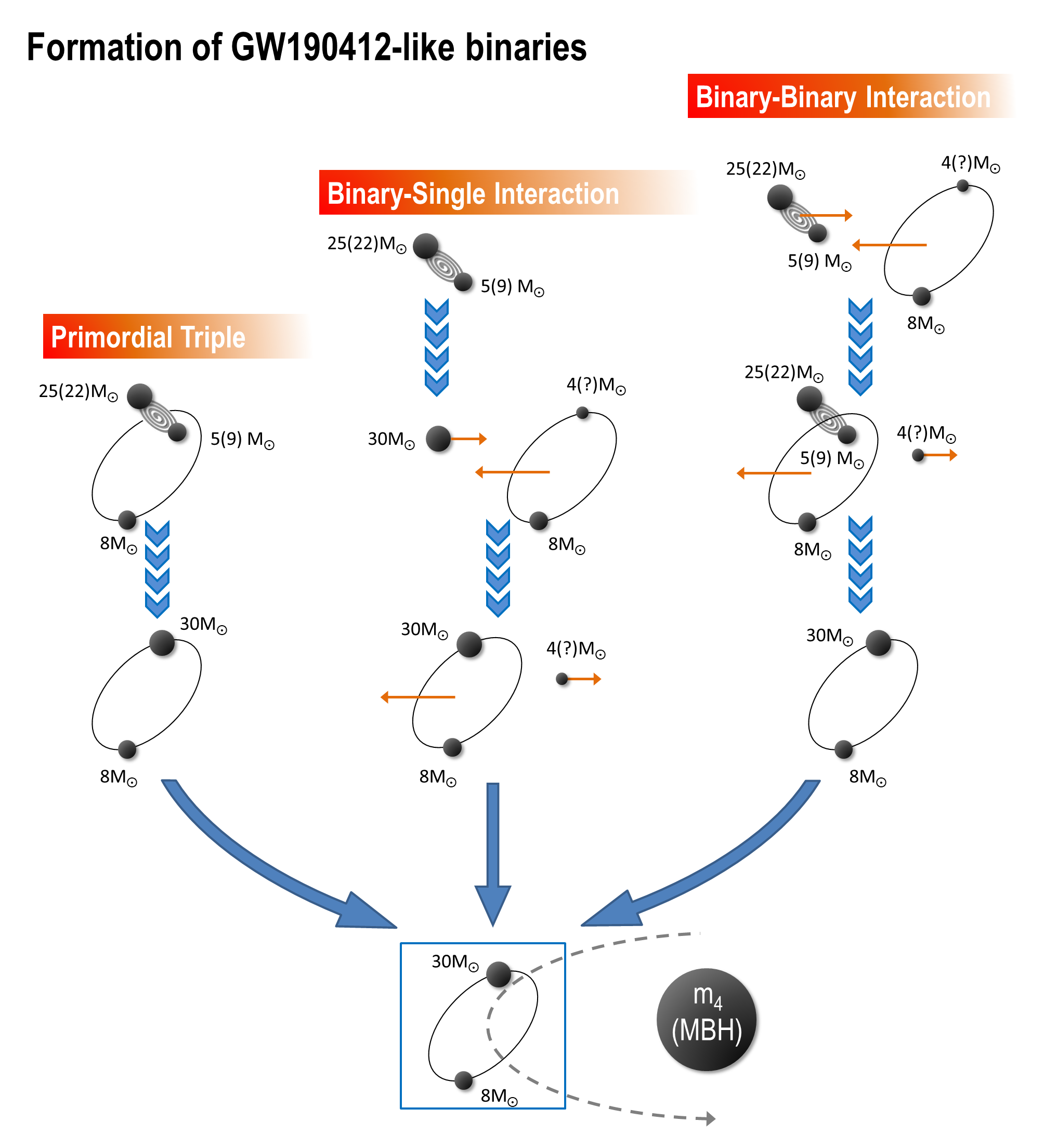}
\end{tabular}
\caption{Different formation pathways of GW190412-like
systems. The final merging BHB can be produced either in a primordial multiple system (left),
or through binary-single interaction (middle) and binary-binary interaction (right).
The final BHB is unlikely to merge by itself, but can be induced to merge by an external body
(likely a massive BH).
We label the possible values of the individual masses based on the analysis in Section \ref{sec 3}
(see also Section \ref{sec 6}).
Two possible sets of the binary progenitor masses are given for the $30M_\odot$ BH.
The number with a question mark implies that the mass is not well constrained.
}
\label{fig:Formation of middle BHB GW190412}
\end{figure}

\begin{figure}
\centering
\begin{tabular}{cccc}
\includegraphics[width=9cm]{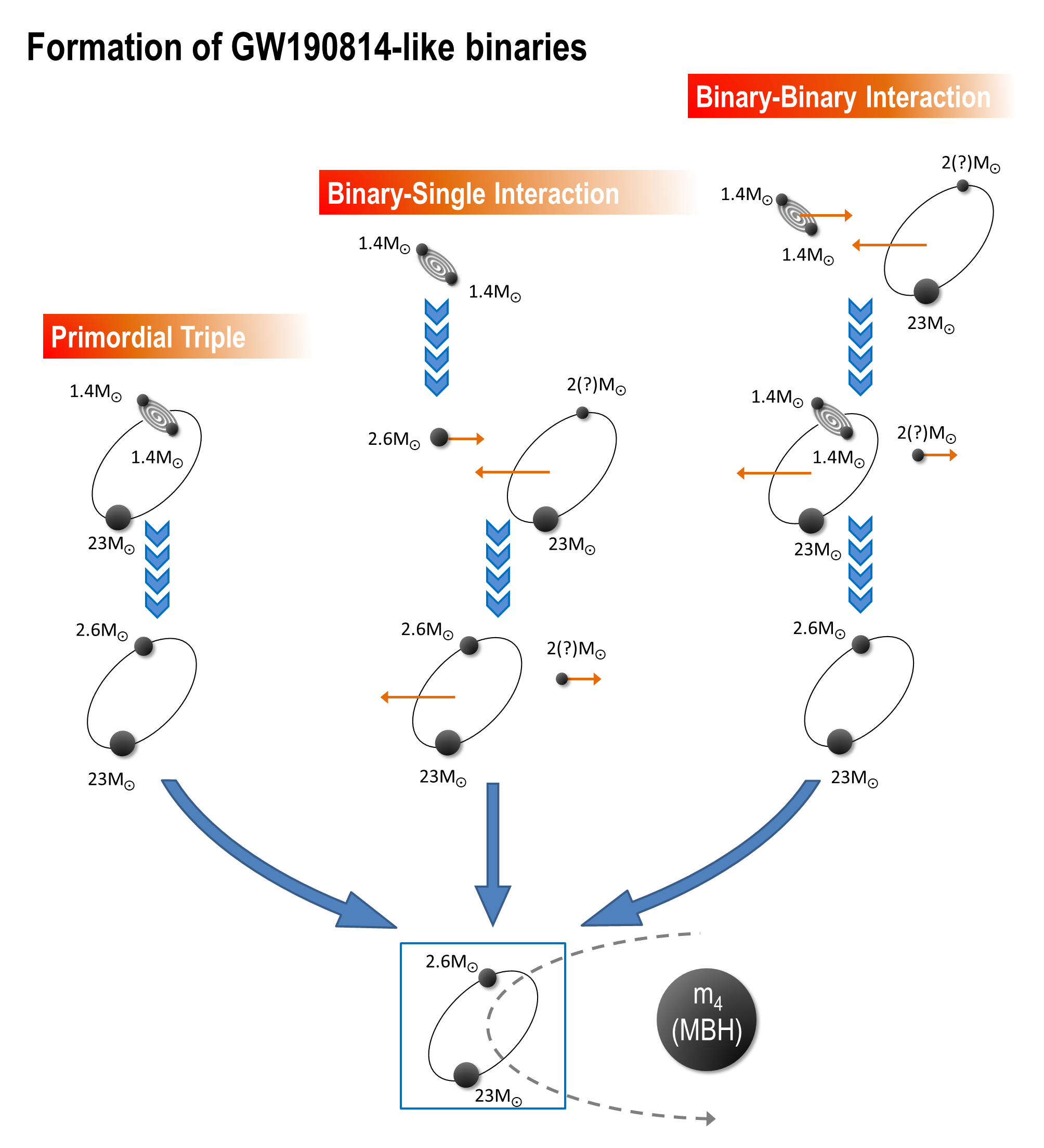}
\end{tabular}
\caption{Similar to Figure \ref{fig:Formation of middle BHB GW190412}, but for GW190814-like systems (see Sections \ref{sec 4} and \ref{sec 6}).
}
\label{fig:Formation of middle BHB GW190814}
\end{figure}

\begin{figure}
\centering
\begin{tabular}{cccc}
\includegraphics[width=9cm]{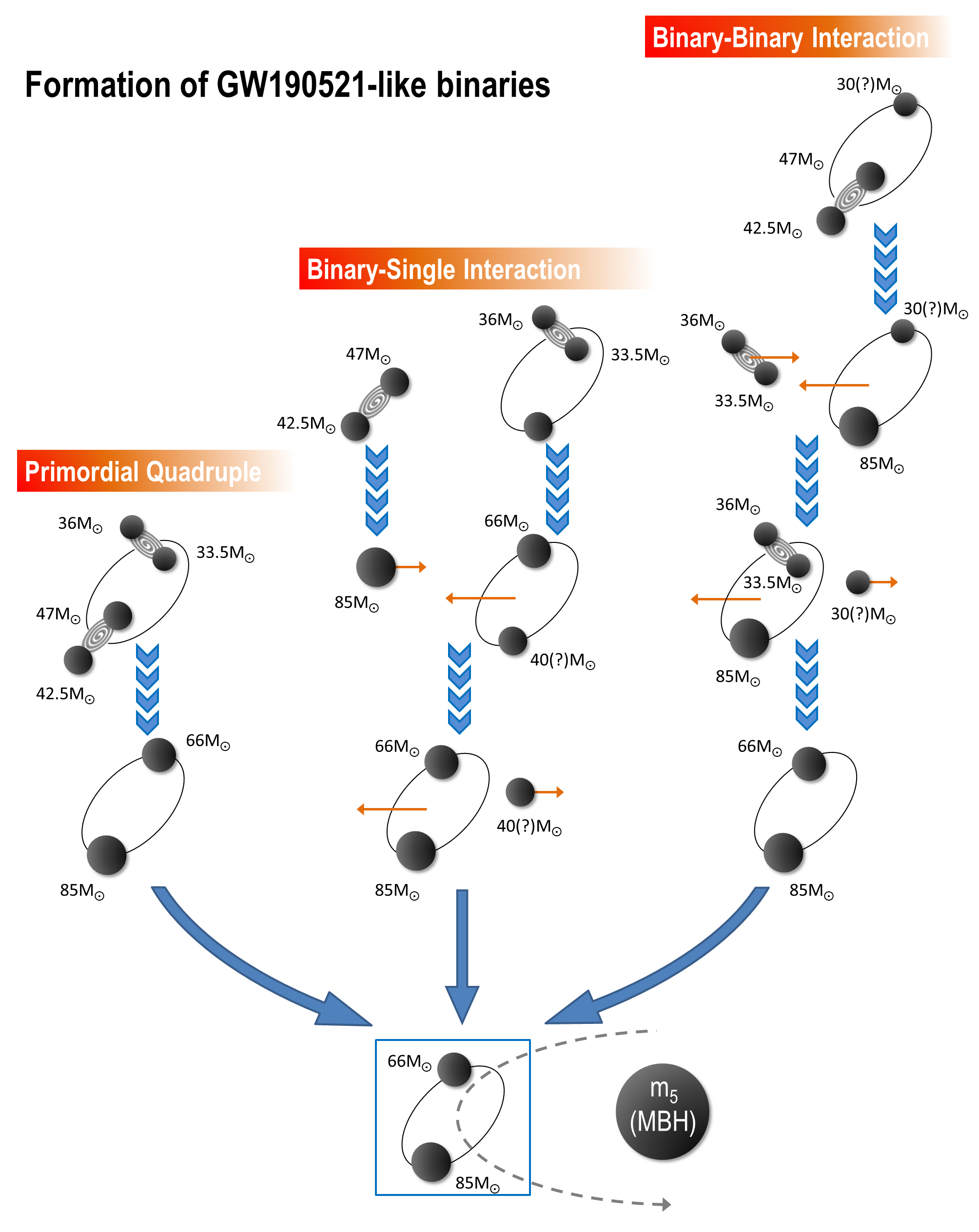}
\end{tabular}
\caption{Similar to Figure \ref{fig:Formation of middle BHB GW190412}, but for GW190521-like systems (see Sections \ref{sec 5} and \ref{sec 6}).
Other possible dynamical pathways involving singe-triple interaction and binary-triple interaction are not shown.
}
\label{fig:Formation of middle BHB GW190521}
\end{figure}

The unexpected features of these three O3 events
may be naturally explained if one or both components are the remnants of the previous BH or NS mergers.
This is generally termed ``hierarchical mergers" \citep[e.g.,][]{GW190521 ApJL},
but exactly how successive mergers occur and with what frequencies are not clear.
In this paper, we study hierarchical mergers involving binaries and multiples,
where the multiples could be either ``primordial" or formed dynamically in dense stellar clusters
(see Figures \ref{fig:Formation of middle BHB GW190412}-\ref{fig:Formation of middle BHB GW190521}).
In particular, we consider the ($30M_\odot$) primary in GW190412 to be a merger product ---
this would explain its large observed spin. We suggest the secondary ($2.6M_\odot$) in GW190814
to be produced by the merger of two canonical ($1.3-1.4M_\odot$) NSs.
We assume both massive components in GW190521 to be the products of first-generation mergers --- this
would explain their large observed spins. Regardless of the detailed evolutionary pathways,
it is likely the final BHBs cannot merge by themselves because of their wide orbital
separations. Instead,
they undergo mergers induced by a tertiary companion, likely a massive or supermassive BH (MBH, SMBH),
through the Lidov-Kozai mechanism \citep[e.g.,][]{Lidov,Kozai}.
Overall, we envision that through different pathways
(Figures \ref{fig:Formation of middle BHB GW190412}-\ref{fig:Formation of middle BHB GW190521}), a final BHB is assembled,
likely in a dense nuclear star cluster, and the final merger is induced by a MBH or SMBH.
We examine the possibility and constraints that systems like GW190412, GW190814 and GW190521 are produced in multiple systems.

\begin{figure}
\centering
\begin{tabular}{cccc}
\includegraphics[width=7cm]{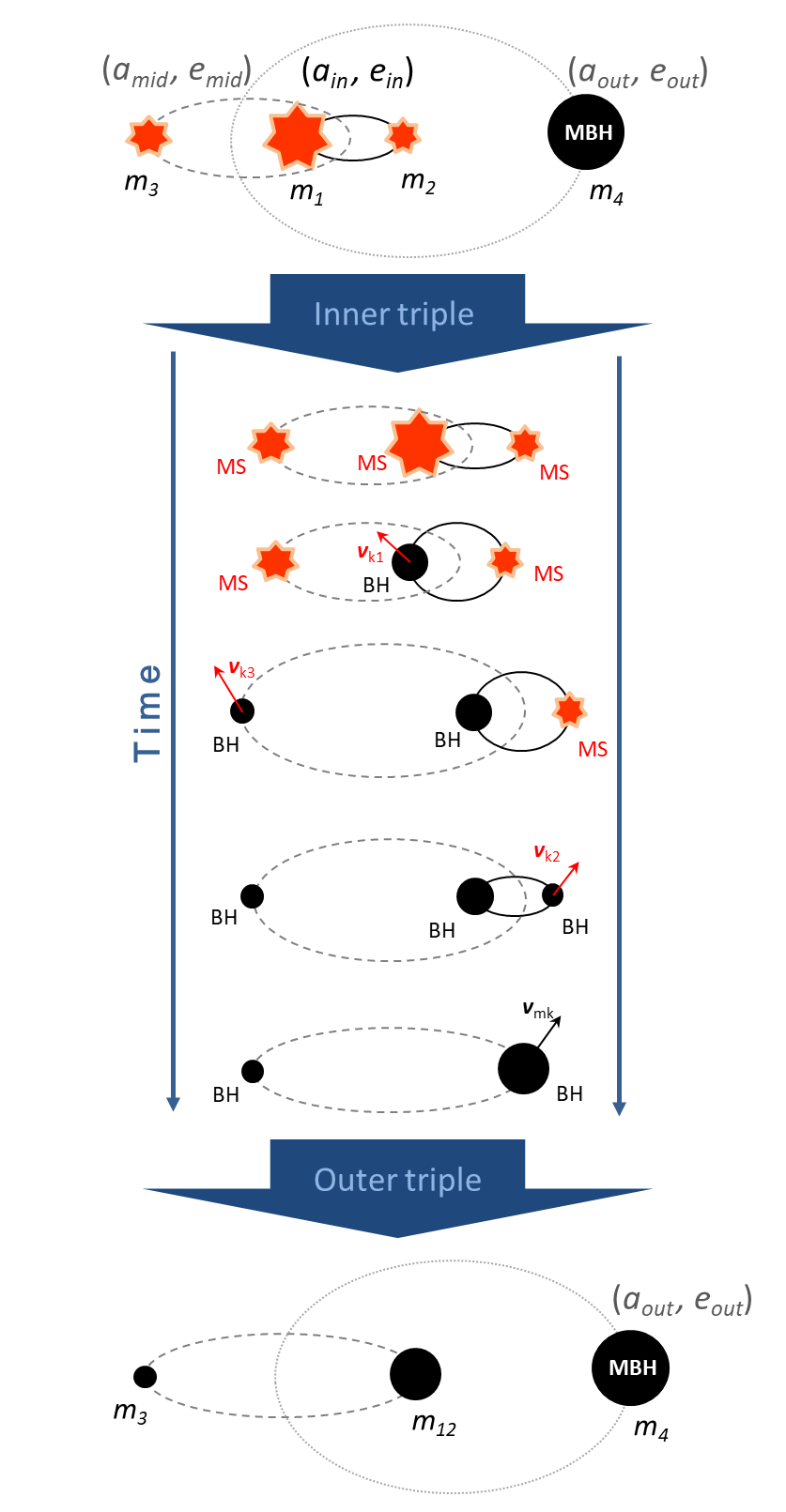}
\end{tabular}
\caption{Detailed evolutionary diagram of a hierarchical quadruple system leading to the formation of GW190412-like systems.
This corresponds to the ``primordial" pathway depicted in Figure \ref{fig:Formation of middle BHB GW190412}.
We consider three nested binaries, where the component masses are $m_1$, $m_2$, $m_3$ and $m_4$
(e.g., the inner triple could reside in a nuclear star cluster surrounding a massive BH $m_4$).
The semi-major axes and eccentricities are denoted by $a_\IN$, $a_\MID$, $a_\OUT$ and $e_\IN$, $e_\MID$, $e_\OUT$, respectively.
The inner triple stars undergo SN explosions accompanied by
natal kicks and mass losses.
The merger of the inner binary is accompanied by a merger kick due to asymmetric GW emission.
If the middle binary survives all the SNs and kicks, it continues the ``final" BHB and can merge due to LK oscillations with the aid of
an external body $m_4$ (a massive or supermassive BH).
}
\label{fig:configuration}
\end{figure}

A specific scenario
for the formation of merging BHBs relies on ``primordial" triples or quadruples
(the leftmost pathway in Figures \ref{fig:Formation of middle BHB GW190412}-\ref{fig:Formation of middle BHB GW190521}).
We study this ``primordial" multiple scenario in details for each of the three LIGO/VIRGO O3 systems.
Figure \ref{fig:configuration} illustrates the key physical processes, using GW190412 as an example.
We consider a hierarchical quadruple system, consisting of three nested binaries.
The innermost binary has masses $m_1$, $m_2$, and moves around $m_3$ forming the middle binary.
This triple system also orbits around the forth body with mass $m_4$, constituting the outer binary.
While $m_1$, $m_2$, $m_3$ have ``stellar" masses
($\sim1M_\odot$ to tens of $M_\odot$), we consider a wide range of possible $m_4$, from stellar mass BH to SMBH.
To produce GW190412, each star in the inner triple undergos stellar evolution and eventually collapses into a BH, possibly accompanied by
a natal kick and sudden mass loss during the supernova (SN) explosion.
The innermost BHB is assumed to merge by itself, as in the standard isolated binary evolution channel,
and the merger remnant receives a merger kick, with the kick magnitude depending on the mass ratio.
Under appropriate conditions (e.g. the kick velocity is not too high),
this newly formed BH ($m_{12}$) may remain bound to $m_3$ in the ``middle" binary to constitute a triple system together with $m_4$
(see Figure \ref{fig:configuration}).
We envision that, with the aid of tertiary companion ($m_4$),
the middle binary may merge eventually via Lidov-Kozai (LK) oscillations
when the mutual inclination between the middle and outer binaries is sufficiently high.
A similar process applies to GW190814, where the secondary component comes from the
merger of two NSs in the innermost binary. For GW190521,
we consider a quadruple system orbiting around an external companion ($m_5$),
where two binaries in the quadruple merge individually, leaving behind a BHB and the tertiary ($m_5$).

Most previous works on BH binary mergers in multiple systems consider specific numerical
examples or carry out some kinds of
``population synthesis" \citep[i.e.,][]{Liu APJ,Liu APJ 2,Fragione 2019a,Hamers 2020} with various assumptions about the progenitor populations.
However, using the results from \citet{Liu APJ},
all essential aspects of the leading-order LK-induced mergers, including the merger ``window" and merger fraction, can be understood analytically.
We summarize and extend these analytical results (including new fitting  formulae)
in this paper (see Section \ref{sec 2 2}). This allows us to constrain the property of the outer binary
($m_4$ or $m_5$ and the semimajor axis $a_\OUT$)
without performing extensive population calculations.

Our paper is organized as follows.
In Section \ref{sec 2},
we present the key physical ingredients for studying BHB mergers in multiple systems.
Section \ref{sec 2 1} describes the effects of natal kicks (associated with NS and BH formation in a SN explosion) and merger kicks on hierarchical systems; these are relevant for the ``primordial" multiple scenario of
Figures \ref{fig:Formation of middle BHB GW190412}-\ref{fig:Formation of middle BHB GW190521}. In Section \ref{sec 2 2}, we
derive the general analytical expressions for the merger fraction and perturbation strength in
tertiary-induced mergers of BHBs; these are relevant for both ``primordial" multiple and ``dynamical" multiple scenarios of
Figures \ref{fig:Formation of middle BHB GW190412}-\ref{fig:Formation of middle BHB GW190521}.
In Section \ref{sec 3}, we explore the formation of GW190412 in the ``primordial" multiple scenario, where the system
experiences natal kicks and mergers in sequence. Based on the distributions of post-kick orbital parameters of the ``final" BHBs,
we constrain the parameter space of the external perturber (the MBH) required to effectively induce mergers of the BHBs.
In Sections \ref{sec 4} and \ref{sec 5}, we examine the formation of GW190814 and GW190521, respectively, using the similar approach.
In Section \ref{sec 6}, we consider the dynamical formation pathways of the
final BHBs (involving single-binary and binary-binary scatterings;
see Figures \ref{fig:Formation of middle BHB GW190412}-\ref{fig:Formation of middle BHB GW190521}).
We discuss our results in Section \ref{sec 7} and summarize our main findings in Section \ref{sec 8}.

\section{Method: Physical Processes in Multiples and in Tertiary-induced Mergers}
\label{sec 2}

We present our method using the evolutionary scenario depicted in Figure \ref{fig:configuration};
this is appropriate for the production of GW190412-like events (see Section \ref{sec 3}).
However, with small adaption, this can be applied to GW190814 (see Section \ref{sec 4})
and GW190521 (Section \ref{sec 5}).

\subsection{Supernova, Natal kick and Merger kick in Triple Systems}
\label{sec 2 1}

In our ``primordial" scenario for the formation of merging BHBs
(see leftmost pathway in Figures \ref{fig:Formation of middle BHB GW190412}-\ref{fig:Formation of middle BHB GW190521}),
the orbital parameters of the final BHB depend on the previous stellar/binary evolutionary history.
Consider the inner triple system depicted in Figure \ref{fig:configuration}.
Since the orbital parameters may change due to SN explosion, we denote the pre-kick and post-kick orbital parameters using the
superscripts ``0" and ``k", respectively.
The pre-SN relative velocity and separation distance of the inner binary and
middle binary are denoted by $\textbf{v}_\IN^0$, $\textbf{r}_\IN^0$, $\textbf{v}_\MID^0$, and $\textbf{r}_\MID^0$, respectively.
We have
\ba
&&|\textbf{v}_\IN^0|=\sqrt{Gm_{12}^0\bigg(\frac{2}{|\textbf{r}_\IN^0|}-\frac{1}{a_\IN^0}\bigg)},\label{eq:Vin}\\
&&|\textbf{v}_\MID^0|=\sqrt{Gm_{123}^0\bigg(\frac{2}{|\textbf{r}_\MID^0|}-\frac{1}{a_\MID^0}\bigg)},\label{eq:Vmid}
\ea
and the angular momenta of two orbits are given by
\ba
&&\textbf{L}_\IN^0\equiv\mu_\IN^0 \textbf{h}_\IN^0=\mu_\IN^0 (\textbf{r}_\IN^0 \times\textbf{v}_\IN^0),\\
&&\textbf{L}_\MID^0\equiv\mu_\MID^0 \textbf{h}_\MID^0=\mu_\MID^0 (\textbf{r}_\MID^0 \times\textbf{v}_\MID^0),
\ea
where $\mu_\IN^0\equiv m_1^0m_2^0/m_{12}^0$ (with $m_{12}^0=m_1^0+m_2^0$),
$\mu_\MID^0\equiv m_{12}^0m_3^0/m_{123}^0$ (with $m_{123}^0=m_1^0+m_2^0+m_3^0$) are the reduced mass for the
inner binary and middle binary, and $\textbf{h}_\IN^0$ and $\textbf{h}_\MID^0$ are the specific angular momenta.

As a star explodes in a SN, its suffers a mass loss $\Delta m$ and expediences a kick.
We assume that the velocity of the natal kick ($v_\mathrm{k}$) is drawn from a Maxwellian
distribution
\be\label{eq:Vkick}
p(v_\mathrm{k})\varpropto v_\mathrm{k}^2 e^{-v_\mathrm{k}^2/\sigma^2}
\ee
with a velocity dispersion $\sigma$.
We also assume the kick direction is isotropically distributed.
The natal kick on the first generation BH is highly uncertain, ranging from $0~\mathrm{km~s^{-1}}$ to $\sim100~\mathrm{km~s^{-1}}$
\citep[e.g.,][]{Repetto 2015,Mandel 2016}.
We will consider both extreme values  (0 and $100~\mathrm{km~s^{-1}}$ in our study.
On the other hand, the kick velocity on a newly formed NS (relevant to GW190814) is well constrained and we adopt
$\sigma=260~\mathrm{km~s^{-1}}$ \citep[e.g.,][]{Hobbs 2005}.

We assume the first SN explosion takes place
on the primary star in the inner binary,
and the newly born BH loses $10\%$ of the mass due to neutrino emission.
The mass of the remnant becomes $m_1^k=m_1^0-\Delta m$.
The natal kick takes place instantaneously compared to the orbital period.
Thus, the relative velocity of the
inner binary is $\textbf{v}_\IN^k=\textbf{v}_\IN^0+\textbf{v}_\mathrm{k1}$
(where $|\textbf{v}_\mathrm{k1}|=v_\mathrm{k1}$; see Equation \ref{eq:Vkick}).
Since $\textbf{r}_\IN^k=\textbf{r}_\IN^0$,
the post-SN semimajor axis is given by
\be\label{eq:a SN}
a_\IN^k=\bigg(\frac{2}{|\textbf{r}_\IN^k|}-\frac{|\textbf{v}_\IN^k|^2}{G(m_1^k+m_2^0)}\bigg)^{-1},
\ee
and the post-SN eccentricity can be obtained from the Laplace-Runge-Lenz vector
\be\label{eq:e SN}
\textbf{e}_\IN^k=\frac{1}{G(m_1^k+m_2^0)}(\textbf{v}_\IN^k\times\textbf{h}_\IN^k)-\frac{\textbf{r}_\IN^k}{|\textbf{r}_\IN^k|},
\ee
where $\textbf{h}_\IN^k=\textbf{r}_\IN^k\times\textbf{v}_\IN^k$.

Note that the SN on $m_1^0$ also imparts a kick on the center of mass (CM) of the inner binary,
which can change the orbital parameters of the middle and outer orbits.
If we define the velocity vector of individual body as $\textbf{v}_{1,2,3}^0$, respectively, we have
$\textbf{v}_\MID^0=\textbf{v}_3^0-\textbf{v}_\mathrm{12 cm}^0$, where
$\textbf{v}_\mathrm{12 cm}^0\equiv(m_1^0\textbf{v}_1^0+m_2^0\textbf{v}_2^0)/m_{12}^0$ is the velocity of the CM of the inner binary.
As $m_1^0$ experiences SN, $\textbf{v}_1^k=\textbf{v}_1^0+\textbf{v}_\mathrm{k1}$,
the post-SN relative velocity of the middle binary becomes \citep[e.g.,][]{Pijloo 2012}
\ba\label{eq:V3n}
&&\textbf{v}_\MID^k=\textbf{v}_3^0-\textbf{v}_\mathrm{12 cm}^k\nonumber\\
&&~~~~~~=\textbf{v}_\MID^0-\frac{m_2^0(m_1^k-m_1^0)}{m_{12}^0
(m_1^k+m_2^0)}\textbf{v}_\IN^0-\frac{m_1^k}{m_1^k+m_2^0}\textbf{v}_\mathrm{k1},
\ea
and the specific angular momentum of the middle binary is given by $\textbf{h}_\MID^k=\textbf{r}_\MID^k\times\textbf{v}_\MID^k$.
The post-SN $a_\MID^k$ and $\textbf{e}_\MID^k$ can be evaluated by replacing
$\textbf{r}_\IN^k\rightarrow \textbf{r}_\MID^k$, $\textbf{v}_\IN^k\rightarrow \textbf{v}_\MID^k$,
$\textbf{h}_\IN^k\rightarrow \textbf{h}_\MID^k$ and $(m_1^k+m_2^0)\rightarrow (m_1^k+m_2^0+m_3^0)$
in Equations (\ref{eq:a SN})-(\ref{eq:e SN}).

The orientations of the angular momenta of two (inner and middle) orbits can change as a consequence of the natal kick.
Taking the middle binary as an example, the angle between the pre-kick and post-kick angular momentum can be computed
by
\be\label{eq:delta Imid}
\Delta I_\MID =\mathrm{arccos}(\hat{\textbf{h}}^0_{\MID}\cdot\hat{\textbf{h}}_\MID^k),
\ee
where $\hat{\textbf{h}}$ is the unit vector.
Note that if the system goes through multiple kicks, the
final angular momentum orientation may change a lot with respective to the original one.

Similar prescriptions can be applied to the cases where the SN explosion takes place on $m_2$ and $m_3$.

Once the inner BH binary merges, the remnant ($m_{12}$) will receive a merger kick.
Assuming the two BHs ($m_1$ and $m_2$) have negligible spins ($\chi_1=\chi_2\ll 1$),
the kick velocity on $m_{12}$ is given by the fitting formula
\citep[e.g.,][]{Lousto 2010}
\be\label{eq:Vmk}
V_\mathrm{mk}=1.2\times10^4\mathrm{km~s^{-1}}\bigg[\frac{\eta^2(1-m_2/m_1)}{1+m_2/m_1}(1-0.93\eta)\bigg],
\ee
where $\eta\equiv(m_2/m_1)/(1+m_2/m_1)^2$.
This merger kick is along the random direction in the inner orbital plane.
The response of the middle binary due to this kick is analogous to the change in the inner binary that has experienced a natal kick
on one component.
The GW emission in the BHB merger is accompanied by energy (``mass") loss, given by
\be\label{eq:GW mass loss}
\frac{\delta m}{m_1+m_2}=0.057\eta+0.445\eta^2+0.522\eta^3.
\ee
The final remnant spin is given by
\be\label{eq:chi 12}
\chi_{12}=\frac{3.46\eta-4.34\eta^2+1.69\eta^3}{(1-0.06\eta-0.44\eta^2-0.52\eta^3)^2}.
\ee

\subsection{Tertiary-Induced Mergers: Analytical Results}
\label{sec 2 2}

\begin{figure*}
\centering
\begin{tabular}{cccc}
\includegraphics[width=17.8cm]{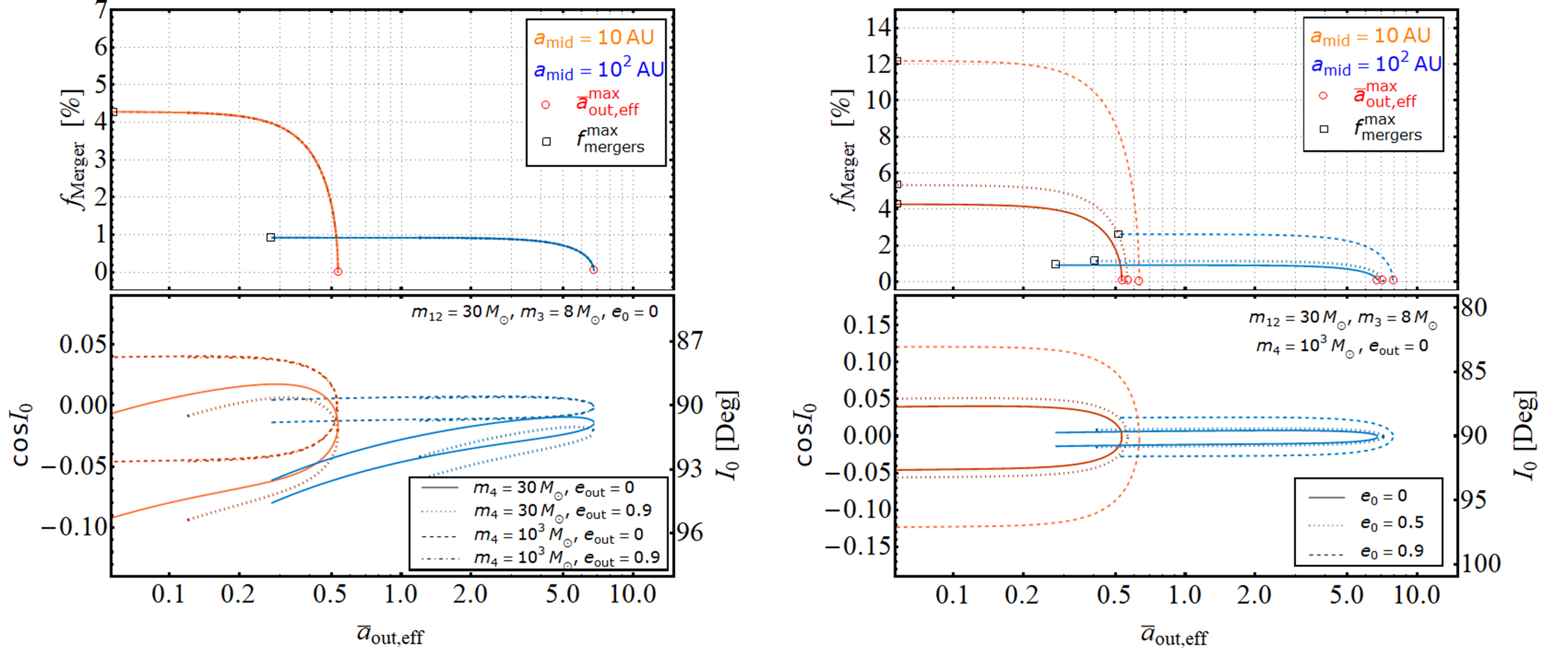}
\end{tabular}
\caption{
Merger fractions and merger windows (see Equation \ref{eq:merger fraction})
as a function of $\bar{a}_{\OUT,\eff}$ (see Equation \ref{eq:aout bar}) for LK-induced mergers.
In this sample plot, we consider the induced merger of the middle binary (with component masses $m_{12}=30M_\odot$ and $m_3=8M_\odot$)
by a tertiary $m_4$.
In the left panels, the middle binary has an initially circular orbit (i.e., initial $e_\MID=e_0=0$,
and two different initial $a_\MID$'s)
and we choose different tertiary companions as labeled.
In the right panels, we fix the tertiary companion mass and eccentricity ($m_4=10^3M_\odot$, $e_\OUT=0$)
but vary the initial eccentricity $e_0$ of the middle binary (as labeled).
These results are obtained analytically using Equations (\ref{eq:EMAX}), (\ref{eq:fitting formula}) and (\ref{eq:merger fraction}).
Each curve terminates on the left at the instability limit.
The maximum value of $\bar{a}_{\OUT,\eff}$ to have a merger is denoted by $\bar{a}_{\OUT,\eff}^\m$ (Equation \ref{eq:fitting a out}),
and the maximum value of $f_\merger$ (which occurs at small $\bar{a}_{\OUT,\eff}$)
is denoted by $f_\merger^\m$ (Equation \ref{eq:fitting f merger}).
}
\label{fig:merger window}
\end{figure*}

Regardless how the ``final" BHB (the middle binary in Figure \ref{fig:configuration}) is produced
(either in the ``primordial" pathway or in one of the ``interaction" pathways, see
Figures \ref{fig:Formation of middle BHB GW190412}-\ref{fig:Formation of middle BHB GW190521}), it would have too large an orbital separation to
merge by itself in most situations. Instead, the middle binary (with mass $m_{12}$ and $m_3$)
can be driven to merge due to the gravitational perturbation from the external body ($m_4$, MBH) that moves on an inclined
outer orbit relative to the orbit of the middle binary.
The LK mechanism induces oscillations in the eccentricity
and inclination of the middle binary on the timescale
\be\label{eq:LK timescale}
T_\lk=\frac{1}{n_\MID}\frac{m_{12}+m_3}{m_4}\bigg(\frac{a_{\OUT,\eff}}{a_\MID}\bigg)^3,
\ee
where $n_\MID=[G (m_{12}+m_3)/a_\MID^3]^{1/2}$ is the mean motion of the middle binary,
and
\be
a_{\OUT,\eff}\equiv a_\OUT\sqrt{1-e^2_\OUT}
\ee
is the effective outer binary semi-major axis.

During LK oscillations, the short-range force effects (such as GR-induced apsidal precession)
play a crucial role in limiting the maximum eccentricity $e_\m$ of the middle binary
\citep[e.g.,][]{Fabrycky 2007}.
In the absence of energy dissipation, the evolution of the
triple is governed by two conservation laws:
the total orbital angular momentum and
the total energy of the system.
An analytical expression for $e_\m$ for general hierarchical triples
(arbitrary masses and eccentricities) can be
obtained in the double-averaged secular approximation if the disturbing potential
is truncated to the quadrupole order
(i.e., when the octupole-order perturbation is negligible; see below).
Using the method of \citet{Liu et al 2015} \citep[see also][]{Anderson et al 2016,Anderson et al 2017}, we find
\ba\label{eq:EMAX}
&&\frac{3}{8}\Bigg\{j_\mi^2-j_0^2+(5-4j_\mi^2)\\
&&\Bigg[1-\frac{\Big((j_\mi^2-j_0^2)\eta-2j_0\cos I_0\Big)^2}{4j_\mi^2}\Bigg]\nonumber\\
&&-(1+4e_0^2-5e_0^2\cos^2\omega_0)\sin^2I_0\Bigg\}+\varepsilon_\gr \left(j_0^{-1}-j_\mi^{-1}\right)=0\nonumber,
\ea
where $e_0$, $I_0$ and $\omega_0$ are the initial eccentricity
\footnote{We denote the initial values of parameters for LK oscillations using the subscript ``0".
Note that the superscript ``0" refers to the pre-SN parameters (see Section \ref{sec 2 2}).},
inclination and longitude of the periapse of the middle binary, respectively,
and we have defined
$j_\mi\equiv\sqrt{1-e_\m^2}$, $j_0\equiv\sqrt{1-e_0^2}$,
$\eta\equiv|\textbf{L}_\MID|/|\textbf{L}_\OUT|$ (at $e_0=0$),
and
\be
\varepsilon_\gr=\frac{3G(m_{12}+m_3)^2a_{\OUT,\eff}^3}{c^2a_\MID^4m_4}.
\ee
Note that for $e_0=0$, Equation~(\ref{eq:EMAX}) reduces to Equation (24) of \cite{Anderson et al 2017}.
For the general $\eta$,
the maximum possible $e_\m$ for all values of $I_0$, called
$e_\li$, is given by (assuming $\omega_0=0$)
\ba\label{eq:ELIM}
&&\frac{3}{8}\Bigg[5-2j_0^2-3j_\li^2+\frac{\eta^2}{4}\bigg(\frac{4}{5}j_\li^2-1\bigg)
\frac{(j_\li^2-j_0^2)^2}{j_\li^2-1}\Bigg]\nonumber\\
&&+\varepsilon_\gr \left(j_0^{-1}-j_\li^{-1}\right)=0,
\ea
where $j_\li\equiv\sqrt{1-e_\li^2}$.

For systems with $m_{12}\neq m_3$ and $e_\OUT\neq 0$, so that
\be
\varepsilon_{\rm oct}\equiv \frac{m_{12}-m_3}{m_{12}+m_3}\bigg(\frac{a_\MID}{a_\OUT}\bigg)
\frac{e_\OUT}{1-e_\OUT^2}
\ee
is non-negligible, the octupole effect may become important \citep[e.g.,][]{Ford,Naoz 2016}.
This tends to widen the inclination window for large eccentricity excitation.
However, the analytic expression for $e_\li$ given by
Equation~(\ref{eq:ELIM}) remains valid even for $\varepsilon_{\rm oct}\neq 0$ \citep[e.g.,][]{Liu et al 2015,Anderson et al 2017}.
In other words, because of the effect of short-range forces due to GR, the maximum eccentricity cannot exceed $e_\li$ even when
the octupole potential is significant.
Note that the stability of triple systems requires \citep[e.g.,][]{Stability condition}
\be\label{eq:stable}
\frac{a_\OUT(1-e_\OUT)}{a_\MID(1+e_\MID)}>\frac{3.7}{Q_\OUT}-\frac{2.2}{1+Q_\OUT}+\frac{1.4}{Q_\IN}\frac{Q_\OUT-1}{Q_\OUT+1},
\ee
where $Q_\IN=[\m(m_{12},m_3)/\mi(m_{12},m_3)]^{1/3}$ and $Q_\OUT=(m_{123}/m_4)^{1/3}$.
In the limit of $m_4\gg m_{123}$, Equation (\ref{eq:stable}) reduces to
$[a_\OUT(1-e_\OUT)/a_\MID(1+e_\MID)]\gtrsim[3.7(m_4/m_{123})^{1/3}]$.
This implies
\ba
&&\varepsilon_{\rm oct}\lesssim\bigg(\frac{m_{12}-m_3}{m_{123}}\bigg)\frac{1}{3.7}\bigg(\frac{m_{123}}{m_4}\bigg)^{1/3}
\frac{e_\OUT}{(1+e_\OUT)(1+e_\MID)}\nonumber\\
&&~~~~~\lesssim1.5\times10^{-3}\bigg(\frac{m_{12}-m_3}{m_{123}}\bigg)\bigg(\frac{10^5m_{123}}{m_4}\bigg)^{1/3},
\ea
where in the second line we have used $e_{\OUT}/(1+e_{\OUT})\leqslant1/2$ and set $e_\MID\simeq1$
(since merger requires high eccentricity).
Therefore, $\varepsilon_{\rm oct}\lesssim0.001$ when $m_4/m_{123}\gtrsim10^5$.
For such systems, the $e$-excitation window is very similar to the quadruple result \citep[e.g.,][]{Liu et al 2015,Diego 2016,Liu SMBH}.
Only a small fraction of mergers are influenced by the octuple effect.

In tertiary-induced mergers, when the octupole effect is negligible,
the merger time can be well approximated by \citep[e.g.,][]{Wen 2003,Thompson 2011,Liu APJ}
\be\label{eq:fitting formula origin}
T_\mathrm{m}\simeq T_\mathrm{m,0}(1-e_\m^2)^3,
\ee
where
\be
T_{\mathrm{m},0}\equiv\frac{5c^5 a_{\MID,0}^4}{256 G^3 (m_{12}+m_3)^2 \mu_\MID}
\ee
is the merger time due to GW emission
for an isolated circular BHB, and $e_{\mathrm{max}}$ is from Equation (\ref{eq:EMAX})
\footnote{
Note that Equation (\ref{eq:fitting formula origin}) only applies in the quadruple approximation,
which is valid for $m_4\gg m_{123}$.
\cite{Hoang 2017} considered BHBs around a SMBH, and found numerical examples where
a fraction of BHBs has the merger time shorter than predicted by Equation (\ref{eq:fitting formula origin}).
This is because their examples refer to BHBs very close to the SMBH, and the systems are near the instability limit.
In any case, only a small fraction of mergers are influenced by the octuple effect.
}.
Using Equation (\ref{eq:fitting formula origin}), we
can define the ``merger eccentricity" $e_\mathrm{m}$ via
\be\label{eq:fitting formula}
T_{\mathrm{m},0}(1-e_\mathrm{m}^2)^3=T_\mathrm{crit}.
\ee
Thus, only systems with $e_\m\gtrsim e_\mathrm{m}$ can have the merger time $T_\mathrm{m}$ less than $T_\mathrm{crit}$.
We typically set $T_\mathrm{crit}=1.4\times10^{10}$ yrs in this paper.
However, for triple systems located in dense star clusters,
such timescale decreases due to the effects from the surrounding stars (see Section \ref{sec 7 2}).
Note that our numerical calculations show that Equation (\ref{eq:fitting formula}) is approximately valid as long as $e_0\lesssim0.9$.
Combining Equations (\ref{eq:EMAX}) and (\ref{eq:fitting formula}),
the merger window (bounded by the critical inclination angles $I_{0,\merger}^\pm$) can be obtained
that allows the middle binary to attain $e_\mathrm{m}$ and merge within $T_\mathrm{crit}$.
If we assume the orbital orientation of tertiary companion is distributed isotropically, the merger fraction
of the BHB is given by \citep[see Section 3 in][]{Liu APJ}
\be\label{eq:merger fraction}
f_\merger=\frac{1}{2}\bigg|\cos I_{0,\merger}^+-\cos I_{0,\merger}^-\bigg|.
\ee

Since the eccentricity excitation depends on $m_4$ (the MBH), $a_\OUT$, $e_\OUT$ through the ratio
$m_4/(a_\OUT\sqrt{1-e_\OUT})^3$ in the quadrupole approximation,
we introduce the dimensionless scaled semi-major axis
\be\label{eq:aout bar}
\bar{a}_{\OUT,\eff}=\bigg(\frac{a_\OUT\sqrt{1-e_\OUT^2}}{1000\au}\bigg)\bigg(\frac{m_4}{10M_\odot}\bigg)^{-1/3}
\ee
to characterize the ``strength" of the external perturber.

Figure \ref{fig:merger window} presents the results of the merger fraction and merger window
for BHB ($m_{12}=30M_\odot$ and $m_3=8M_\odot$) as a function of $\bar{a}_{\OUT,\eff}$,
obtained using Equations (\ref{eq:EMAX}), (\ref{eq:fitting formula})-(\ref{eq:aout bar}) and
$T_\mathrm{crit}=1.4\times10^{10}$ yrs.
All the systems shown here satisfy the stability criterion for the triples \citep[e.g.,][]{Stability condition}.
We choose two different initial semi-major axes ($a_\MID=10\au$ and $100\au$).
In the left panels,
for each BHB, we fix the initial eccentricity of the middle BHB to zero and
consider a variety of outer binaries (different $m_4$ and $e_\OUT$, as labeled).
We find that, for a given $a_\MID$, different $m_4$ and $e_\OUT$ (with the same $\bar{a}_{\OUT,\eff}$) affect the position of merger window
(i.e. the range of $\cos I_0$) but not the value of $f_\merger$.
We see that $f_\merger$ is approximately constant until $\bar{a}_{\OUT,\eff}$ is close to a maximum
value ($\bar{a}_{\OUT,\eff}^\m$), beyond which $f_\merger$ quickly drops to zero.
Also, the merger window and fraction have strong dependence on the initial semi-major axis ($a_\MID$).
This is because for small $a_\MID$, the induced eccentricity in the LK oscillations
does not have to be too large to produce mergers within $T_\mathrm{crit}$.
In the right panels, we fix the outer binary and consider the effect of different $e_0$ (the initial value of $e_\MID$).
We find that the merger window and fraction can be increased by a factor of a few if the BHB is initialized as an eccentric orbit.
In addition, we see that the range of $\bar{a}_{\OUT,\eff}$ producing merger is different for different separation of BHBs.

\begin{figure}
\centering
\begin{tabular}{cccc}
\includegraphics[width=7cm]{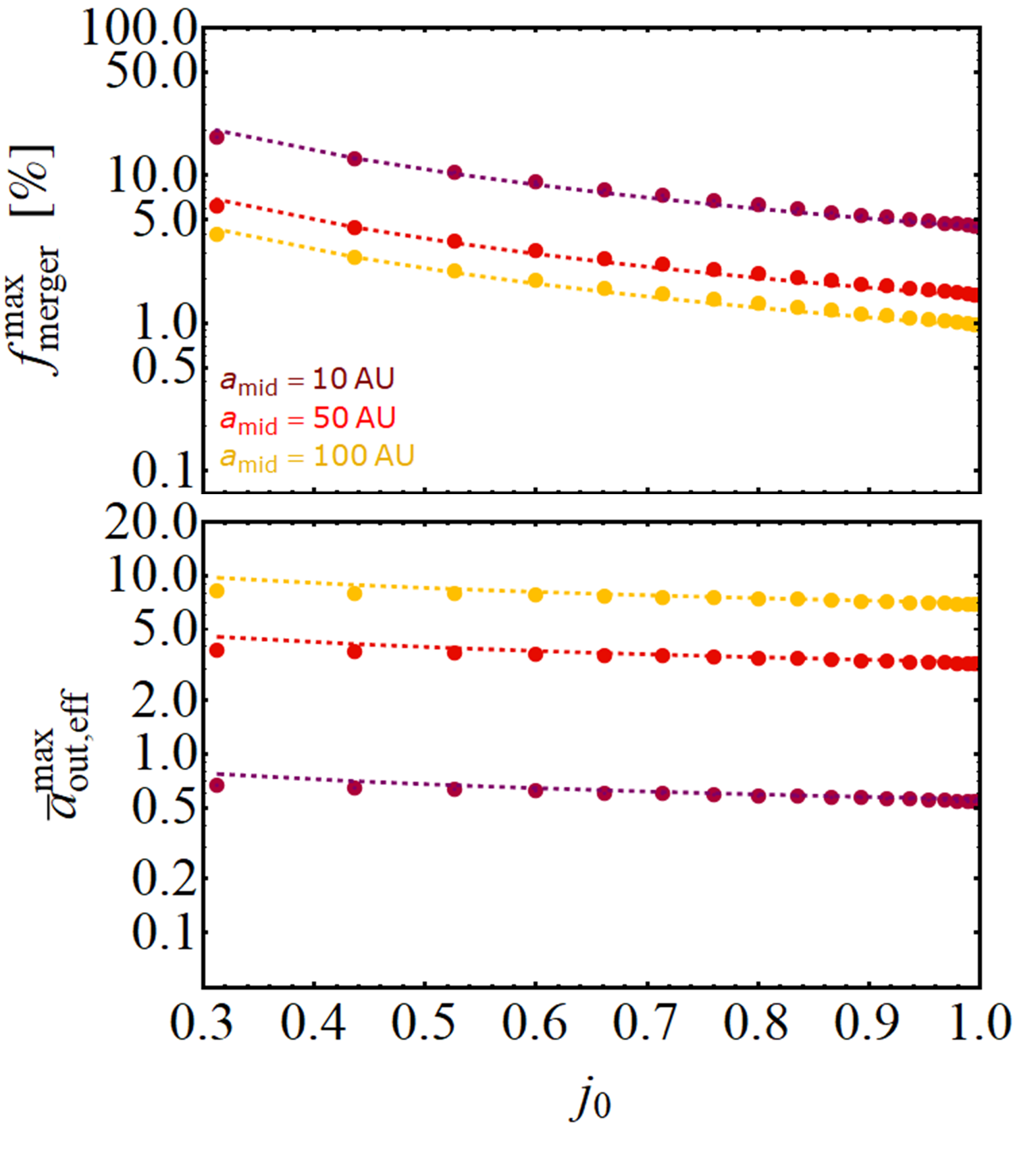}
\end{tabular}
\caption{
The maximum values of $f_\merger$ and $\bar{a}_{\OUT,\eff}$ (see Figure \ref{fig:merger window}) as a function of $j_0=\sqrt{1-e_0^2}$ of
the BHB (the middle binary).
The points are obtained analytically using Equations (\ref{eq:EMAX}), (\ref{eq:fitting formula}) and (\ref{eq:merger fraction}).
The dashed lines are given by the fitting formulae (\ref{eq:fitting f merger}) and (\ref{eq:fitting a out}).
}
\label{fig:maximum merger fraction}
\end{figure}

In \citet{Liu APJ} (see their Equations 53-54), we showed that the
maximum merger fraction $f_\merger^\m$ (i.e., the maximum value of $f_\merger$ for a given initial $a_\MID$)
and $\bar{a}_{\OUT,\eff}^\m$ (the maximum value of $\bar{a}_{\OUT,\eff}$ beyond which $f_\merger$ drops to zero)
can be characterized by two simple fitting formulas when the BHB is initialized in a circular orbit (initial $e_\MID=0$).
Here, we extend the fittings to cover BHBs with different initial eccentricities.
From Equation (\ref{eq:fitting formula}),
we see that the critical eccentricity $e_\mathrm{m}$ required for merger within time $T_\mathrm{crit}$ depends on
$(\mu_\MID T_\mathrm{crit})(m_{123}/a_\MID^2)^2$ for initial $e_\MID=0$, where $m_{123}=m_{12}+m_3$ and $\mu_\MID=m_{12} m_3/m_{123}$.
From Equation (\ref{eq:EMAX}) we see that
the critical inclinations for a given $e_\m=e_\mathrm{m}$
depend on $\varepsilon_\gr$, or the combination $(m_{123}/a_\MID^2)^2(a_{\OUT,\eff}^3/m_4)$, and $j_0=\sqrt{1-e_0^2}$.
Thus the merger fraction $f_\merger$ depends on $m_{12}$, $m_3$, $a_\MID$, $e_0$ and $T_\mathrm{crit}$ only through
$j_0$, $m_{123}/a_\MID^2]$ and $\mu_\MID T_\mathrm{crit}$.
Based on these, we expect that
$f_\merger^\m\propto j_0^\alpha(a_\MID/m_{123}^{0.5})^{-0.67}(\mu_\MID T_\mathrm{crit})^{0.16}$
and $\bar{a}_{\OUT,\eff}^\m\propto j_0^\beta (a_\MID/m_{123}^{0.5})^{1.1}(\mu_\MID T_\mathrm{crit})^{0.06}$,
where $\alpha$, $\beta$ are fitting parameters.
Figure \ref{fig:maximum merger fraction} shows that $\alpha=-1.3$ and $\beta=-0.28$ provide good fit. Thus we have
\be\label{eq:fitting f merger}
\begin{split}
f_\merger^\m\simeq&~4.35\%~(1-e_0^2)^{-0.65}\bigg(\frac{\mu_\MID}{M_\odot}\frac{T_\mathrm{crit}}{10^{10}\mathrm{yrs}}\bigg)^{0.16}\\
&\times\bigg[\bigg(\frac{a_\MID}{\au}\bigg)\bigg(\frac{m_{12}+m_3}{M_\odot}\bigg)^{-0.5}\bigg]^{-0.67},
\end{split}
\ee
and
\be\label{eq:fitting a out}
\begin{split}
\bar{a}_{\OUT,\eff}^\m\simeq&~0.29~(1-e_0^2)^{-0.14}\bigg(\frac{\mu_\MID}{M_\odot}\frac{T_\mathrm{crit}}{10^{10}\mathrm{yrs}}\bigg)^{0.06}\\
&\times\bigg[\bigg(\frac{a_\MID}{\au}\bigg)\bigg(\frac{m_{12}+m_3}{M_\odot}\bigg)^{-0.5}\bigg]^{1.1}.
\end{split}
\ee
These fitting formulae are valid for any type of LK-induced BH mergers in the quadrupole order
(which is valid when the tertiary is a MBH; see \citet{Liu APJ} for how octuple effects can increase the merger fraction).

\section{Formation of GW190412}
\label{sec 3}

In this section, we examine the formation of GW190412 in the ``primordial multiple " scenario
(see Figure \ref{fig:configuration} and the leftmost pathway of Figure \ref{fig:Formation of middle BHB GW190412}).
The formation of the ``final" binary due to dynamical interactions is discussed in Section \ref{sec 6}.

\subsection{Properties of the Inner Binary}
\label{sec 3 1}

In our scenario (Figures \ref{fig:Formation of middle BHB GW190412} and \ref{fig:configuration}),
GW190412 is a tertiary-induced merger, where the primary component ($m_{12}$) in the middle binary
is itself the merger product of an inner binary ($m_1$ and $m_2$).
Since the spin magnitude of the ``second generation (2G) BH" ($\chi_{12}$) in this source has already been determined
\citep[$\chi_{12}=0.43^{+0.16}_{-0.26}$;][]{GW190412},
the mass ratio of two progenitors can be constrained if the 1G BHs have negligible spins
--- this is reasonable if $m_1$ and $m_2$ are a natural product of stellar/binary evolution
\citep[e.g.,][]{Fuller 2019a,Fuller 2019b}.

\begin{figure}
\centering
\begin{tabular}{cccc}
\includegraphics[width=7.3cm]{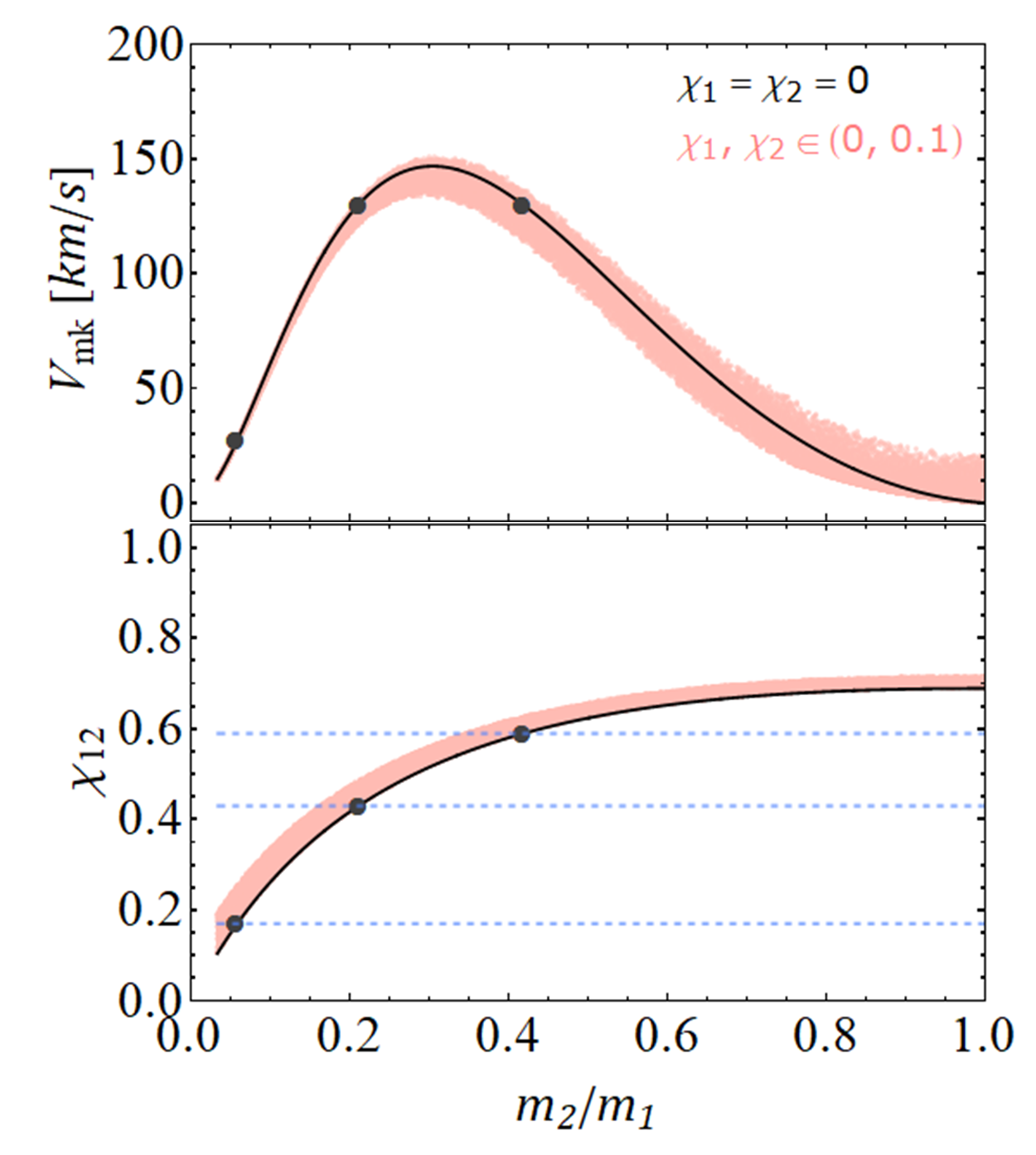}
\end{tabular}
\caption{The merger kick (upper panel) and remnant spin (lower panel) as a function of binary mass ratio.
The solid lines are from Equations (\ref{eq:Vmk}) and (\ref{eq:chi 12}) for $\chi_1=\chi_2=0$,
and the red bands are obtained
assuming progenitors are slowly rotating with $\chi_1, \chi_2\lesssim0.1$ (see the analytic fits in \citet{Lousto 2010}).
The dashed-blue lines in the lower panel indicate the range of the observed spin magnitude of the primary component of GW190412,
i.e., $\chi_{12}=0.43^{+0.16}_{-0.26}$.
From this, the mass ratio of the progenitors and the kick velocity can be constrained. The black dots indicate the three cases shown in
Table \ref{tab:Parameters of Progenitors}.
}
\label{fig:m1m2}
\end{figure}

In Figure \ref{fig:m1m2}, we use Equations (\ref{eq:Vmk}), (\ref{eq:chi 12}) and
plot the merger kick ($V_\mathrm{mk}$) and spin ($\chi_{12}$) as a function of $m_2/m_1$,
assuming no natal spin for the two 1G BHs (see black curves).
For reference, if each 1G BH has a small but finite spin ($\chi_1, \chi_2\lesssim0.1$)
and the associated spin orientation is aligned with the orbital angular momentum,
the values of $V_\mathrm{mk}$ and $\chi_{12}$ (obtained by the analytic fits in \citet{Lousto 2010})
would exhibit only small spread (see the red band in Figure \ref{fig:m1m2}).
Using the observed value of $\chi_{12}=0.43^{+0.16}_{-0.26}$,
the mass ratio is constrained to be $m_2/m_1=0.211^{+0.208}_{-0.153}$.
For concreteness, we adopt $m_{12}=30M_\odot$, and introduce
three cases to cover the possible range for the parameters of the progenitors of $m_{12}$
(see Table \ref{tab:Parameters of Progenitors}).
Note that we have included the radiated energy from GW on the component masses (see Equation \ref{eq:GW mass loss});
thus $m_{12}=m_1+m_2-\delta m$.

\subsection{Constrain the post-kick orbit of BHB (the middle binary)}
\label{sec 3 2}

\begin{table}
 \centering
 \begin{minipage}{80mm}
  \caption{Possible parameters of the progenitors of the $30M_\odot$ primary in GW190412, following the constrains from Figure \ref{fig:m1m2}.
\label{tab:Parameters of Progenitors}
  }
  \begin{tabular}{@{}llrrrrlrlr@{}}
  \hline
    & ~~~~~Case I &~~~~~ Case II    &~~~~~   Case III  \\
 \hline
$\chi_{12}$ & ~~~~~~~0.43 & 0.17~~~ & 0.59~~~~\\
$ m_2/m_1$ & ~~~~~~0.211 & 0.058~~ & 0.419~~~\\
$ m_1, m_2 [M_\odot]$ & ~~~~25.2, 5.3 & 28.5, 1.6& 21.9, 9.2 \\
$ V_\mathrm{mk} [\mathrm{km/s}]$ & ~~~~~~~130 & 27.1~~~ & 129.8 ~~\\
\hline
\end{tabular}
\end{minipage}
\end{table}

Having obtained the possible masses $(m_1, m_2)$ of the inner binary from the GW data
on GW190412, (Section \ref{sec 3 1}), we now constrain what kind of middle binary
($m_{12}=30M_\odot$ around $m_3=8M_\odot$) can survive the three SNs and the merger kick without becoming unbound.
In our fiducial example, we adopt the parameters in Case I (see Table \ref{tab:Parameters of Progenitors}), and assume the natal kick is negligible
($\sigma_\mathrm{BH}=0~\mathrm{km~s^{-1}}$ in Equation \ref{eq:Vkick}).
In our scenario, $m_1$ and $m_2$ merge through the standard binary evolution channel
\footnote{The innermost binary may also merge due to the perturbation from $m_3$ through LK oscillations.
But this process depends on the mutual inclination angle as well as several additional effects on $m_1$ and $m_2$
during the stellar evolution, i.e., tidal effect and mass transfer, which can suppress the eccentricity excitation.
}.
So we choose
$a_\IN^0=0.1$ AU and $e_\IN^0=0$ at the initial time (right before the SN explosion).
These are reasonable in the standard binary evolution model \citep[e.g.,][]{Belczynski 2016},
although a range of the initial values of $a_\IN^0$ are possible.
Our choice of $a_\IN$ only affects $a_\MID$, $e_\MID$ through the stability requirement for the inner triple (see below).
Our strategy for constraining the middle binary is as follows:

\begin{figure}
\centering
\begin{tabular}{cccc}
\includegraphics[width=8cm]{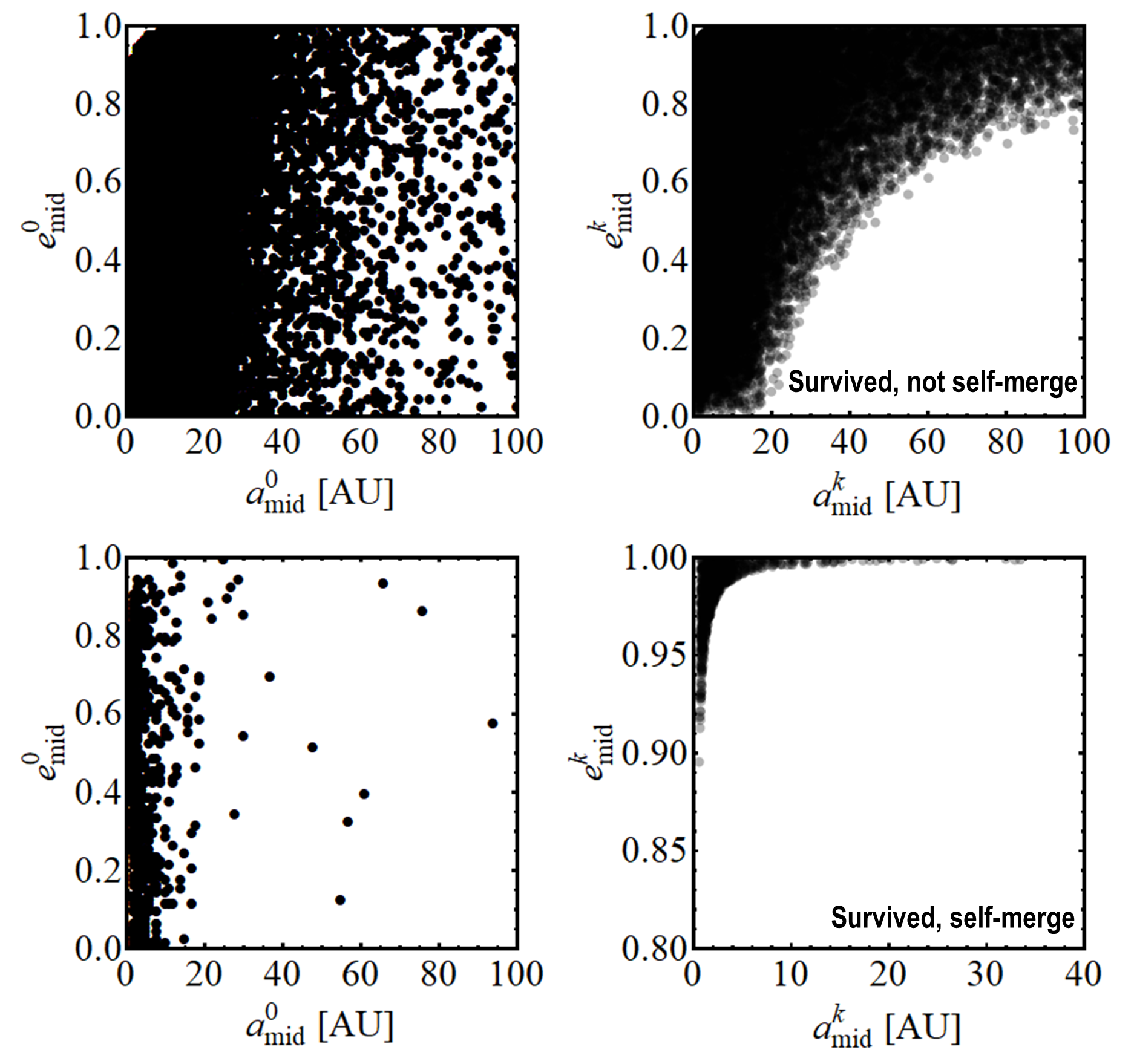}
\end{tabular}
\caption{The distributions of $a_\MID$ and $e_\MID$ for the survived middle binaries that have undergone
four kicks (three natal kicks and one merger kick), where
the right panels show the post-kick systems while the left panels show their pre-kick precentors.
The post-kick middle binaries in the upper right panel cannot merge within the Hubble time by themselves,
while the survived binaries in the lower right panel can.
In our calculation, the pre-kick component masses are chosen to be $m_1^0=28M_\odot$, $m_2^0=5.9M_\odot$ and $m_3^0=8.9M_\odot$
and post-kick masses are
$m_1^k=25.2M_\odot$, $m_2^k=5.3M_\odot$ (Case I in Table \ref{tab:Parameters of Progenitors}) and $m_3^k=8M_\odot$.
Note that the pre-SN mass is larger than the post-SN mass by about $10\%$.
The natal kicks for each newly born BH is assumed to be negligible ($\sigma_\mathrm{BH}=0~\mathrm{km~s^{-1}}$) but the merger kick
is set to be $V_\mathrm{mk}=130~\mathrm{km~s^{-1}}$ (see Table \ref{tab:Parameters of Progenitors}).
}
\label{fig:BHB sigma 0 case I Hubbletime}
\end{figure}

\begin{itemize}
\item We sample a $100\times100$ uniform grid in the plane of $a_\MID^0-e_\MID^0$,
with the pre-SN semi-major axis $a_\MID^0$ ranging from 0 to 100 AU, and eccentricity $e_\MID^0$ ranging from 0 to 1.
This parameter space is further constrained by the stability criterion
\citep[e.g.,][]{Stability condition} for the inner triple system ($m_1, m_2, m_3$).
Note that in principle, the initial middle binaries may have a wider range of $a_\MID^0$ (e.g., to $10^4$ AU or larger in the galactic field).
Such wide binaries ($a_\MID^0\gtrsim 100$ AU) are much less common: e.g., in the galactic field, the typical semi-major axis distribution
is roughly $\propto (a_\MID^0)^{-1.5}$ or steeper \citep[see][]{Moe 2017,El 2018,Tian 2020};
the distribution in dense star cluster is unknown.
Moreover, such wide binaries can be easily destroyed by various kicks (see below).
Since our main goal is to determine the range of $a_\MID$, $e_\MID$ for the survived middle binaries,
it is not important to sample larger values of $a_\MID^0$.

\item For each allowed (i.e. stable) triple system (specified by the values of $a_\MID^0$ and $e_\MID^0$),
we let SN occur on each mass component
and the merger kick on $m_{12}$, following the sequence shown in Figure \ref{fig:configuration}.
Every time a kick happens, we draw a random orbital phase (uniform distribution of the orbital mean anomaly),
and only keep the stable post-kick systems.
In order for $m_1$ and $m_2$ to merge within the Hubble time by themselves,
we also remove systems that have a longer merger time ($\gtrsim1.4\times10^{10}$ yrs) when both $m_1$ and $m_2$ become BHs.

\item Since the changes of the orbital parameters depend on the orientations of the kick velocities
(i.e., natal kick has a random orientation, and merger kick is in the orbital plane),
to cover all possibilities,
we repeat the previous step for 100 times for each ($a_\MID, e_\MID$).
Thus, the statistical features of the survivals can be characterized by the accumulated data.
\end{itemize}

Figure \ref{fig:BHB sigma 0 case I Hubbletime}
shows the distributions of the survived binaries (right panels) and their progenitors (left panels) that have undergone 3 natal kicks
and one merger kick.
We separate the post-kick systems into two categories: the survived middle binaries that cannot
merge over the cosmic time by themselves(upper panels) and those that can (lower panels).
In the upper left panel, we exhibit the pre-kick distributions of $a_\MID^0$ and $e_\MID^0$ for the survived binaries.
We see that only the binaries with sufficiently small $a_\MID^0$ ($\lesssim40\au$)
can survive the SNs and kicks --- binaries with larger $a_\MID^0$ are easily destroyed.
In the upper right panel, we show the post-kick orbital parameters of survived middle binaries in the $a_\MID^k-e_\MID^k$ plane.
We see that these systems tend to have small $a_\MID^k$ and larger $e_\MID^k$.
Note that the kicks can either harden or soften the binary. In the bottom right panel,
we find that some of the post-kick binaries in our simulations can merge within the Hubble time in isolation;
these all have small separations and extreme large orbital eccentricities.
Since the fraction of such systems is quite small ($\sim2\%$), we suggest that an external companion ($m_4$, MBH) is required to
induce the merger of post-kick survived binaries.

\begin{figure*}
\centering
\begin{tabular}{cccc}
\includegraphics[width=17.8cm]{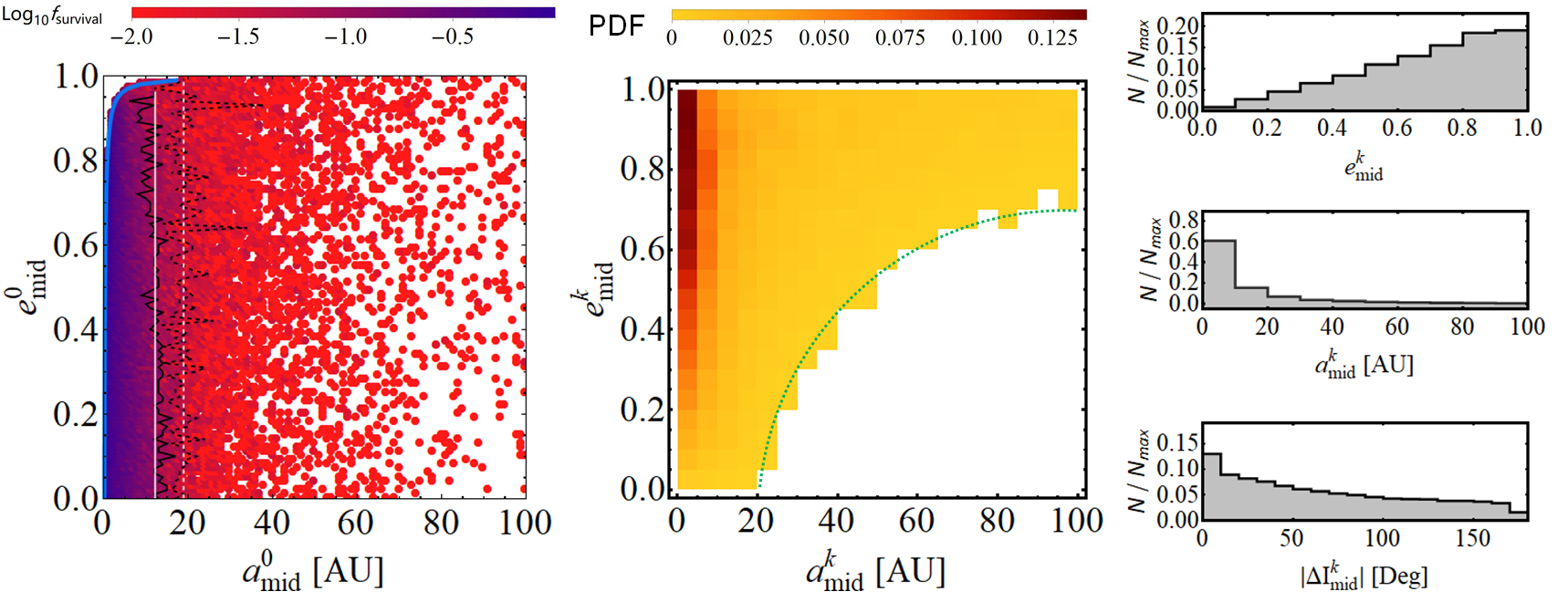}
\end{tabular}
\caption{
Same as Figure \ref{fig:BHB sigma 0 case I Hubbletime}, but combining all survived middle binaries.
The left panel shows the pre-kick systems that have been constrained by the stability criterion
of the inner triple (assuming the pre-kick inner binary has $a_\IN^0=0.1 \au$)
(solid-blue line),
and the survival fractions are color-coded.
The systems with $f_\mathrm{survival}>10\%$ ($5\%$) are enclosed by the black-solid (dashed) line,
and the white (solid and dashed) lines represent the averages.
The middle panel shows the post-kick systems that have survived; the PDF (color coded) denotes the
distributions of $a_\MID^k$ and $e_\MID^k$;
we see that most of the survived systems
cluster around the parameter region with relative small $a_\MID^k$ and large $e_\MID^k$.
The green-dotted line represents an approximate boundary below which no system is produced.
The three rightmost panels show the probability distribution of $e_\MID^k$, $a_\MID^k$ and $|\Delta I_\MID^k|$
(the change in the orientation of the middle binary due to the kicks, see Equation \ref{eq:delta Imid}).}
\label{fig:BHB sigma 0 case I}
\end{figure*}

Figure \ref{fig:BHB sigma 0 case I} shows the similar results as Figure \ref{fig:BHB sigma 0 case I Hubbletime}, but includes all survived binaries.
In the left panel,
the blue line denotes the stability limit of the inner triples.
Each dot implies that there is at least one successful survived binary, and the color represents the surviving possibility (as labeled).
The systems with high survival fraction, where $f_\mathrm{survival}>10\%$ ($5\%$), are enclosed by
solid-black (dashed-black) line.
We also average the values of $a_\MID^0$ on these black (solid and dashed) lines
and highlight them as the solid-white (dashed-white) lines to show the approximate boundaries.
In the middle panel, we show the PDF of all post-kicks survived middle binaries.
We find that the majority of survived systems cluster around the region with $a_\MID^k<10\au$ and $e_\MID^k>0.6$ (see the middle panel).
We also present the distributions of the post-kick orbital parameters in the right panels.
In the bottom rightmost panel, we show the change in the orientation of $\textbf{L}_\MID$ after experiencing 4 kicks
(from the initial $\textbf{L}_\MID^0$).
We see that $|\Delta I_\MID^k|$ has a broad range, indicating the orientation of $\textbf{L}_\MID$ can change significantly.

\begin{figure}
\centering
\begin{tabular}{cccc}
\includegraphics[width=8cm]{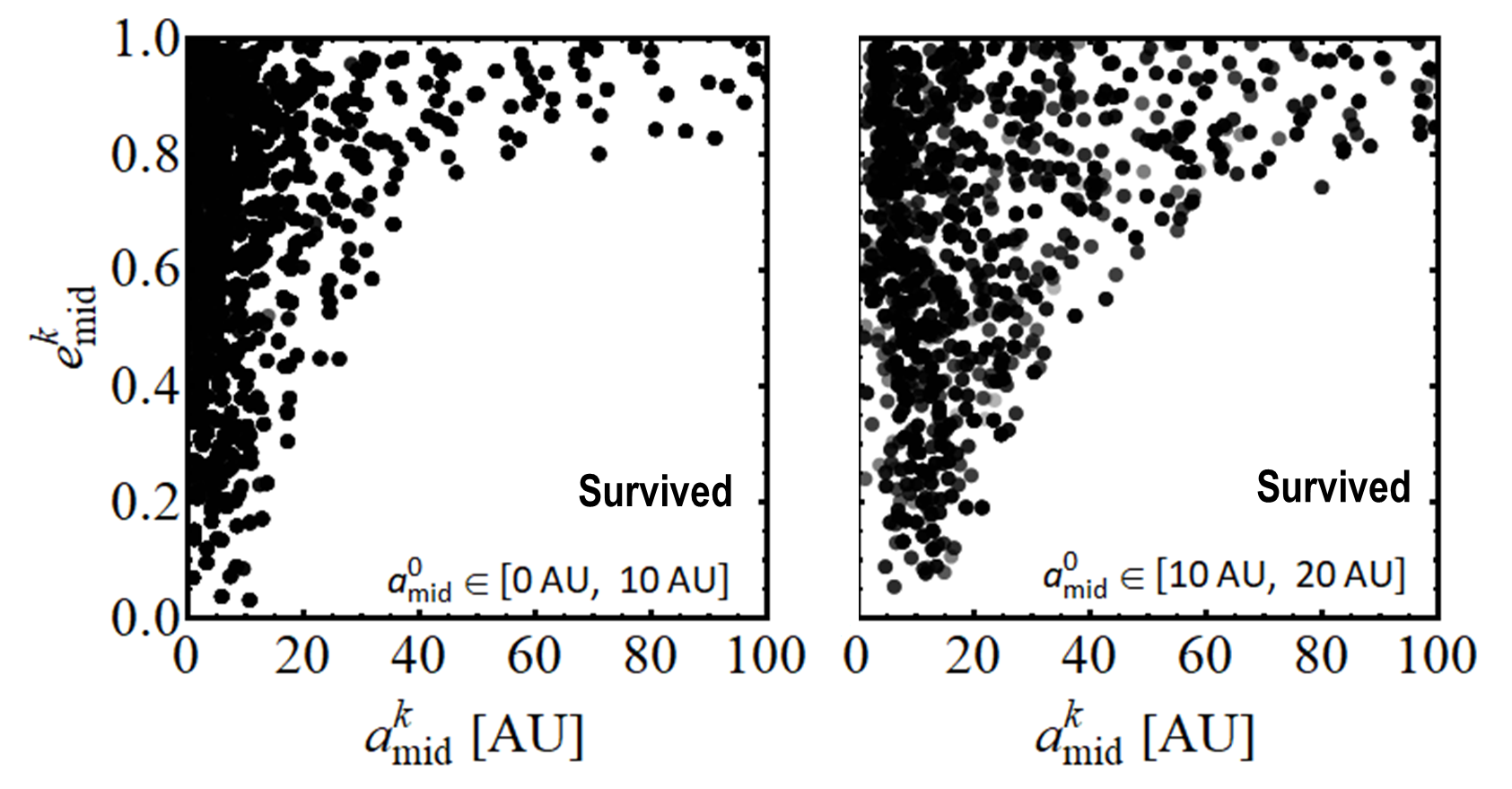}
\end{tabular}
\caption{The survived middle binaries in the $a_\MID^k-e_\MID^k$ plane,
where $a_\MID^0$ is initialized with different ranges (as labeled).
}
\label{fig:BHB sigma 0 case I different a0}
\end{figure}

To examine the dependence of the survived middle binaries on the initial semimajor axis, we
show in Figure \ref{fig:BHB sigma 0 case I different a0} the systems from two ranges of $a_\MID^0$.
We see that the final distributions of $a_\MID^k$ and $e_\MID^k$ have a weak dependence on the initial middle binary parameters.
In another word, the middle binary with specific $a_\MID^k$ and $e_\MID^k$ can be produced by a wide range of ($a_\MID^0$, $e_\MID^0$) values.

\subsection{Constrain the BHB and Tertiary Companion in LK-induced Mergers}
\label{sec 3 3}

Having constrained the range of properties of the middle binary
($m_{12}$ around $m_3$) following three SNe and the first merger
(between $m_1$ and $m_2$ leading to $m_{12}$), we now constrain the property of the tertiary $m_4$ in LK-induced mergers.
Based on the result of Section \ref{sec 2 2}, the maximum merger fraction in LK-induced mergers is only determined by the
properties of the BHB (the middle binary) and is independent of the property of the tertiary companion (see Equation \ref{eq:fitting f merger}).
Thus, we can directly identify what kind of BHB systems may have relatively high merger fraction by using Equation (\ref{eq:fitting f merger}).

\begin{figure}
\centering
\begin{tabular}{cccc}
\includegraphics[width=8.5cm]{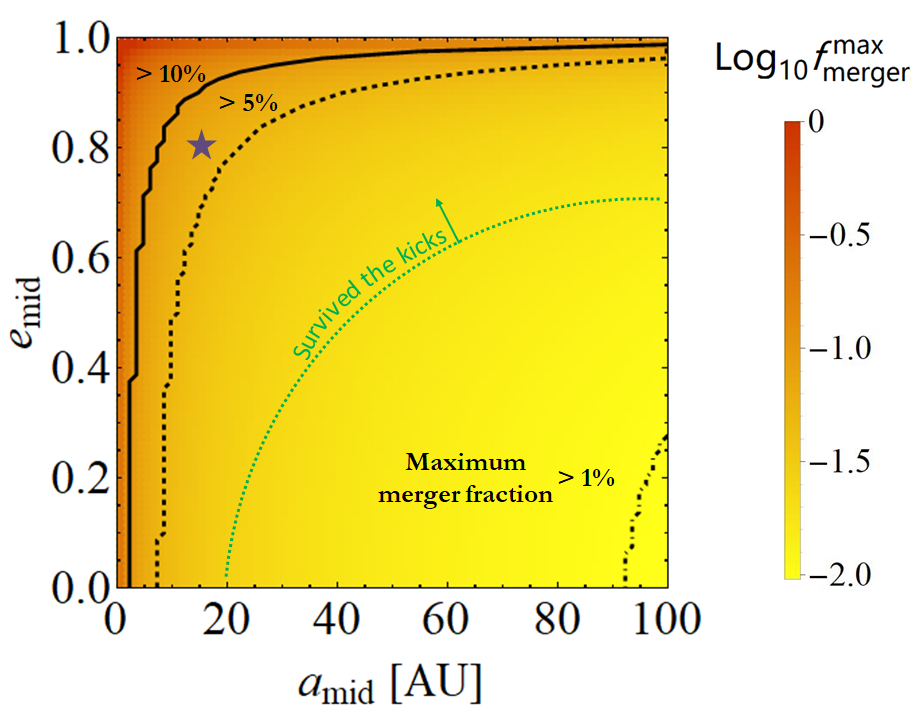}
\end{tabular}
\caption{The maximum merger fraction of BHBs (the middle binaries) induced by the tertiary companion
(Equation \ref{eq:fitting f merger}) in the $a_\MID-e_\MID$ parameter plane
(this is the same as $a_\MID^k=e_\MID^k$ in Figure \ref{fig:BHB sigma 0 case I}).
The binary masses are $m_{12}=30M_\odot$ and $m_3=8M_\odot$ (appropriate for GW190412).
The three black lines (solid, dashed and dot dashed) specify $f_\merger^\m=10\%,5\%$ and $1\%$, respectively.
The green-dotted line is from the middle panel of Figure \ref{fig:BHB sigma 0 case I}, indicating systems that have survived three SNe and kicks.
}
\label{fig:Merger fraction in middle BHB}
\end{figure}

In Figure \ref{fig:Merger fraction in middle BHB}, we show the values of $f_\merger^\m$ in the $a_\MID-e_\MID$ parameter space,
for BHBs with $m_{12}=30M_\odot$ and $m_3=8M_\odot$ (as appropriate for GW190412).
We see that almost all the systems with $a_\MID$ less than about $100\au$ have the merger fraction greater than $1\%$ (dot-dashed line).
Higher fractions can be achieved if the BHB is either sufficiently eccentric, or compact
(see the contour solid-black line and dashed-black line showing $f_\merger^\m=10\%$, $5\%$, respectively).
Since the parameter space of the middle binary is also constrained by SNe and kicks (Section \ref{sec 3 2}),
we consider the systems above the green-dashed line (referring to the survived systems in Figure \ref{fig:BHB sigma 0 case I}).
Thus, all survived systems can potentially
have high merger fractions ($\gtrsim$ a few $\%$) if there is a sufficiently ``strong" tertiary companion.

\begin{figure}
\centering
\begin{tabular}{cccc}
\includegraphics[width=7cm]{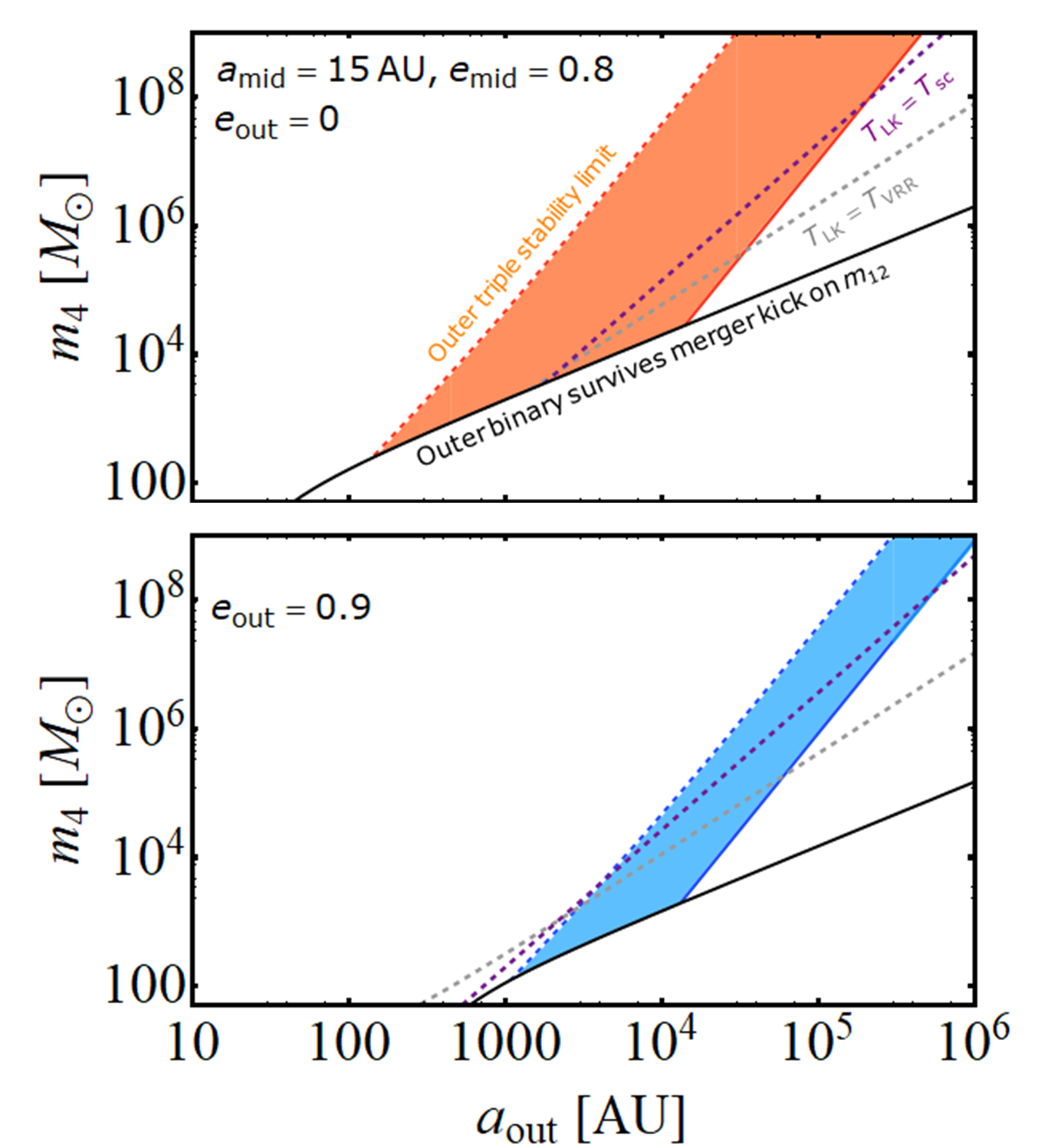}
\end{tabular}
\caption{
Constraints on the mass ($m_4$) and the orbital semimajor axis $a_\OUT$ of the tertiary companion
(MBH) that can lead to LK-induced mergers of the BHB (the middle binary), resembling GW190412.
We consider a representative example of the middle binary (with $a_\MID=15 \au$, $e_\MID=0.8$)
with a relatively high $f_\merger^\m$ (between $5\%$ and $10\%$; see the star in  Figure \ref{fig:Merger fraction in middle BHB}),
and include two values of the outer eccentricity: $e_\OUT=0$ (upper panel) and $e_\OUT=0.9$ (lower panel).
The allowed (shaded) regions are constrained by the stability criterion of the outer triple (orange- and blue-dashed lines), $\bar{a}_{\OUT,\eff}^\m$
(orange- and blue-solid lines) and by requiring that the outer binary remains bound in the presence of the merger kick on $m_{12}$.
The two purple-dashed and gray-dashed lines in each panel indicate the
region where the star cluster potential and vector resonant relaxation effect can enhance the BHB merger rate
(see Equations \ref{eq:VRR timescale} and \ref{eq:cluster potential}).
}
\label{fig:m4 orbit}
\end{figure}

To constrain the parameters of outer binary in our scenario, we pick
one representative system with $a_\MID=15\au$ and $e_\MID=0.8$ (see the star symbol shown in Figure \ref{fig:Merger fraction in middle BHB}).
The constraints on $m_4$ and $a_\OUT$ are shown in Figure \ref{fig:m4 orbit}.
In the upper panel, we assume the outer orbit is circular.
The orange-dashed line is given by the stability criterion \citep[e.g.,][]{Stability condition}, and the orange-solid line
corresponds to the weakest tertiary companion ($\bar{a}_{\OUT,\eff}^\m$) obtained by Equation (\ref{eq:fitting a out}).
We also consider the influence of the merger kick, which happens on $m_{12}$
and affects the outer binary in an indirect way.
This is analogous to the effect on the middle binary due to the natal kick on $m_1$ (see Section \ref{sec 2 1}).
In this case, the orbital velocity of the outer binary changes due to the the merger kick (cf. Equation \ref{eq:V3n}).
By setting the binding energy of the outer binary to zero,
we find the boundary where the outer binary remains bound (black line).
We see that only a sufficiently massive $m_4$ can produce GW190412 like events (color-shaded region in Figure \ref{fig:m4 orbit}).
In the lower panel of Figure \ref{fig:m4 orbit}, we consider the case of $e_\OUT=0.9$.
We see that the allowed parameter region becomes even more restricted.

If we consider a somewhat different middle binary, e.g., $a_\MID=10\au, e_\MID=0.9$,
similar result as Figure \ref{fig:m4 orbit} can be produced, except that the color-shaded region is slightly shifted to the left side.
We therefore conclude that the constraint on $m_4$, $a_\OUT$ as depicted in Figure \ref{fig:m4 orbit} is quite representative.

A general conclusion from Figure \ref{fig:m4 orbit} is that to induce merger of the BHB,
the tertiary companion must be at least a few hundreds $M_\odot$, and falls in the intermediate-mass BH and SMBH regime.

\subsection{The Effect of Different Natal Kick and Merger Kick}
\label{sec 3 4}

\begin{figure*}
\centering
\begin{tabular}{cccc}
\includegraphics[width=17.8cm]{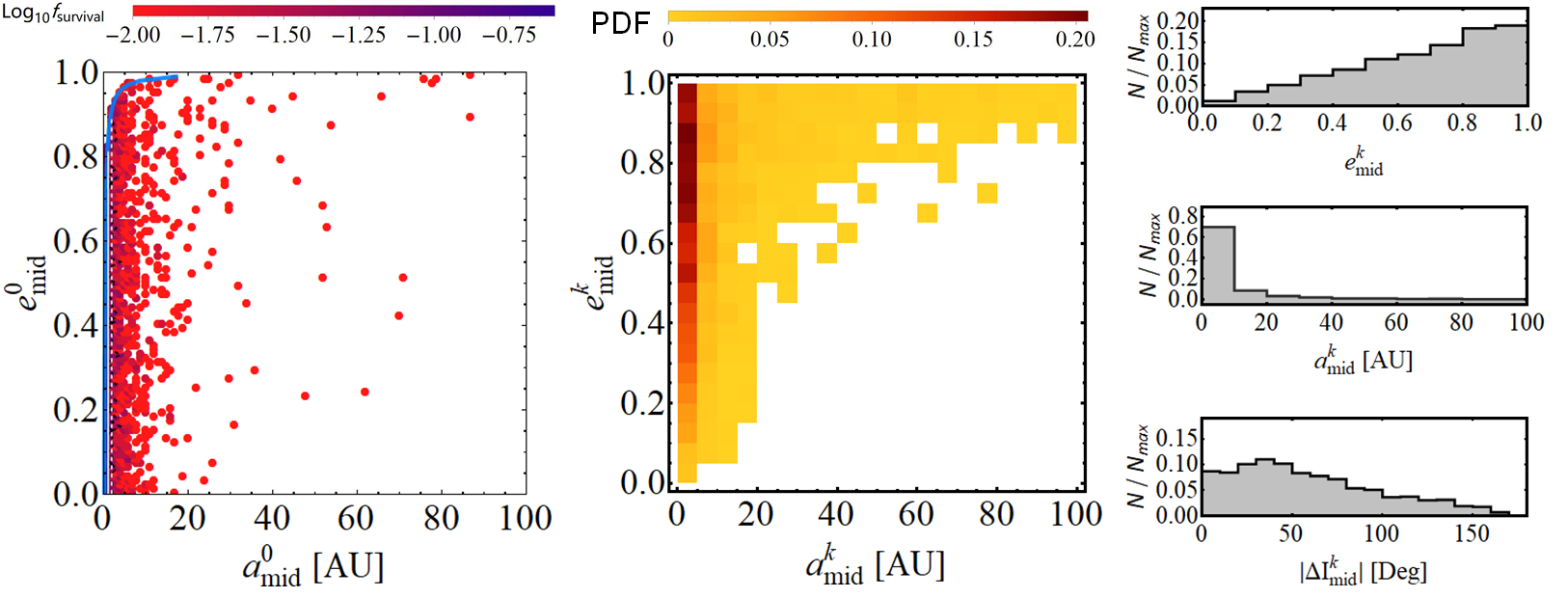}\\
\end{tabular}
\caption{
Similar to Figure \ref{fig:BHB sigma 0 case I}, but with a larger natal kick for BH formation
($\sigma_\mathrm{BH}=100~\mathrm{km~s^{-1}}$).
}
\label{fig:BHB sigma 0 100}
\end{figure*}

In Sections \ref{sec 3 2} and \ref{sec 3 3}, we have assumed that the three stellar-mass BHs all suffered negligible natal kicks.
We now examine how larger natal kicks ($\sigma_\mathrm{BH}=100~\mathrm{km~s^{-1}}$) may affect our constraints.
We carry out the similar analysis as in Section \ref{sec 3 2}, still using the parameters shown in
Case I in Table \ref{tab:Parameters of Progenitors},
but increase the natal kick velocity when SN explosion occurs.
The results are shown in Figure \ref{fig:BHB sigma 0 100}.
In this case, all four kicks are of order of $\sim100~\mathrm{km~s^{-1}}$.
We see that the survival fraction $f_\mathrm{survival}$ is low for all reasonable values of initial $a_\MID^0$.

\begin{figure*}
\centering
\begin{tabular}{cccc}
\includegraphics[width=17.8cm]{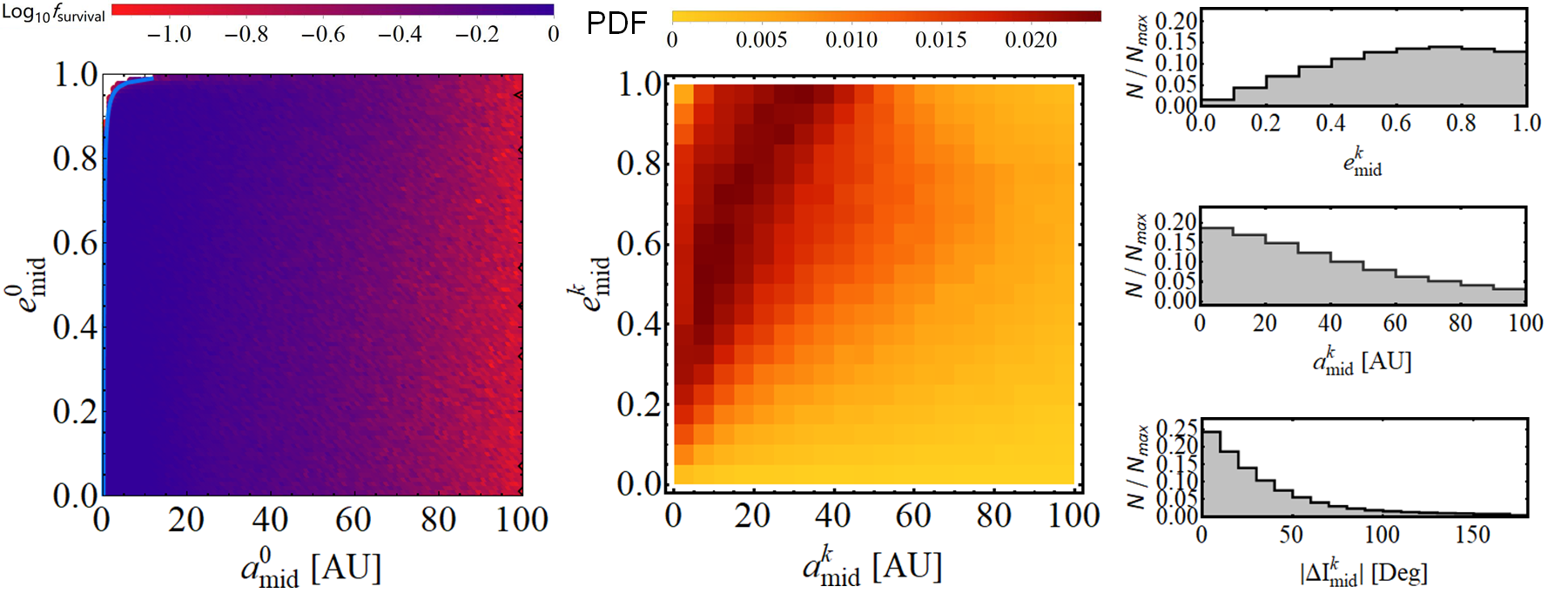}\\
\end{tabular}
\caption{
Same as Figure \ref{fig:BHB sigma 0 case I}, but with small merger kick
of $V_\mathrm{mk}=27.1~\mathrm{km~s^{-1}}$ (see Case II in Table \ref{tab:Parameters of Progenitors}).
}
\label{fig:BHB sigma 0 case II}
\end{figure*}

We also consider the effect of different values of merger kick
(see Figure \ref{fig:BHB sigma 0 case II}). We evolve the systems with the parameters shown in
Case II in Table \ref{tab:Parameters of Progenitors}.
In this case, both the natal kicks and merger kick are negligible.
As a result, almost all the middle binaries we considered can survive with high probabilities, and the survived post-kick $a_\MID^k$ and $e_\MID^k$
cover the whole parameter region (see Figure \ref{fig:BHB sigma 0 case II}).
Of course, in this case, the innermost binary ($m_1$ and $m_2$) has extreme asymmetric masses
(mass ratio $=0.058$), which pose
question about its formation.
Finally, note that the systems from case III in Table \ref{tab:Parameters of Progenitors} have similar behavior as case I,
since the merger kick is almost the same in the two cases.

\section{Formation of GW190814}
\label{sec 4}

\begin{figure}
\centering
\begin{tabular}{cccc}
\includegraphics[width=5cm]{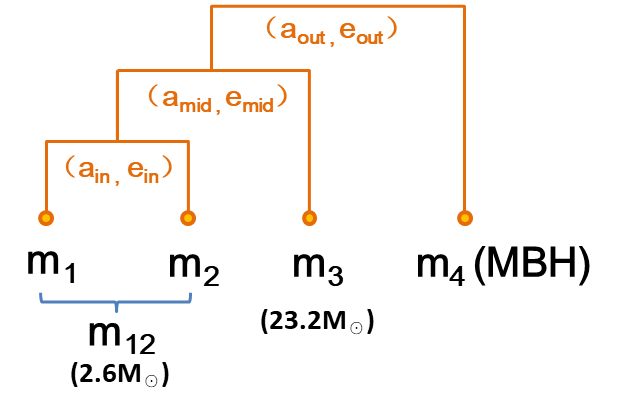}
\end{tabular}
\caption{Schematic diagram of a hierarchical quadruple system for forming GW190814, where
the secondary component (mass $2.6 M_\odot$) is a merger remnant of two NSs.
}
\label{fig:configuration GW190814}
\end{figure}

We now turn our attention to the formation of GW190814
in the ``primordial multiple" scenario (see the leftmost pathway in Figure \ref{fig:Formation of middle BHB GW190814}) .
Different from the event GW190412, where the primary component is a merger product,
we suggest that the secondary component (mass $2.6 M_\odot$) in GW190814 is the remanet from a previous merger of two NSs
(see Figure \ref{fig:configuration GW190814}).
Since the spin parameter of the secondary in GW190814 is not constrained from the GW data,
we cannot constrain the mass ratio ($m_2/m_1$) as in the case of GW190412 (Section \ref{sec 3 1}).
Instead, we assume that the $2.6 M_\odot$ secondary results from the merger of two $1.4 M_\odot$ NSs,
and each NS has evolved from a Helium star of mass $4 M_\odot$.
The SN explosion leads to significant mass ejection, accompanied by a large natal kick ($\sigma_\mathrm{NS}=260~\mathrm{km~s^{-1}}$).

To produce GW190814-like BHBs in our scenario,
we evolve the inner triple systems (see Figure \ref{fig:configuration GW190814})
that undergo three natal kicks and one merger kick.
Different from GW190412, the first SN occurs on $m_3$, which is the most massive component.
We set the natal kick on this BH to zero.
Then, the other two progenitors experience SN explosions and the newly born NSs
receive significant natal kicks. Over time, the binary NSs merge by themselves.
The merger kick received on the remnant is negligible because $m_1=m_2$ (see Figure \ref{fig:m1m2}).

\begin{figure*}
\centering
\begin{tabular}{cccc}
\includegraphics[width=17.8cm]{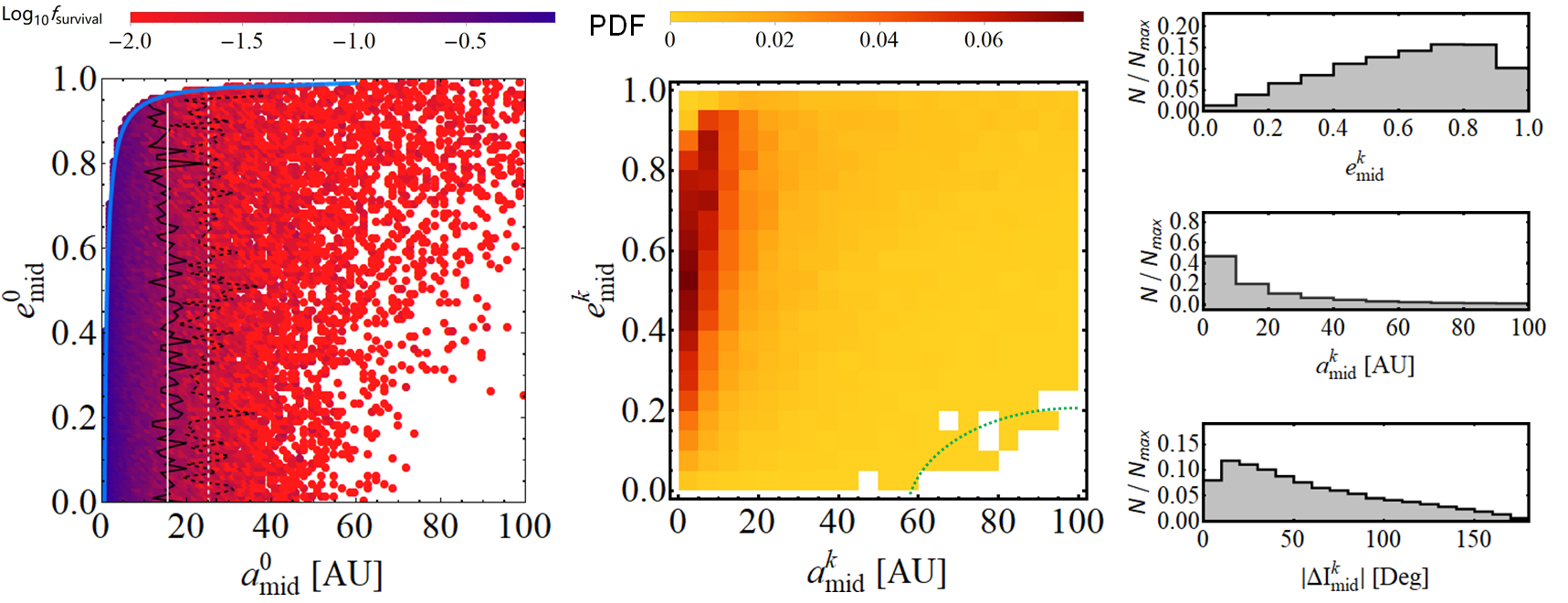}
\end{tabular}
\caption{
Similar to Figure \ref{fig:BHB sigma 0 case I}, but for the systems that lead to GW190814-like binaries.
In this calculation, the pre-SN component masses are $m_1^0=m_2^0=4M_\odot$ and $m_3^0=25.8M_\odot$,
and post-SN masses are
$m_1^k=m_2^k=1.4M_\odot$ and $m_3^k=23.2M_\odot$.
In the inner binary, each newly formed NS receives
a large natal kick ($\sigma_\mathrm{NS}=260\mathrm{km~s^{-1}}$) during the SN explosion. For the post-SN BH, we assume
$\sigma_\mathrm{BH}=0\mathrm{km~s^{-1}}$.
The merger of $m_1^k$ and $m_2^k$ leads to the remnant with mass $m_{12}=2.6M_\odot$ and with negligible merger kick.
}
\label{fig:NSB sigma 0}
\end{figure*}

Following the same strategy as described in Section \ref{sec 3 2}, we obtain the orbital distributions of the survived
middle binaries.
The results are shown in Figure \ref{fig:NSB sigma 0}.
The left panel presents the pre-kick distributions of $a_\MID^0$ and $e_\MID^0$.
The systems enclosed by the solid-blue line are stable,
where we set $a_\IN^0=0.1\au$ and $e_\IN^0=0$ for the inner binary \citep[e.g.,][]{Belczynski 2018}.
Note that unlike the case of GW190412,
the distributions of $a_\MID^0$ and $e_\MID^0$ for the survived systems depend on the $a_\IN^0$ and $e_\IN^0$
sensitively: e.g., if we choose $a_\IN^0=0.14\au$,
the number of survived systems can decrease by over $90\%$.
This results from the fact that a large natal kick on the NS can easily destroy the systems
and some of the remaining NS binaries cannot merge within a Hubble times.
Similar to Figure \ref{fig:BHB sigma 0 case I},
the middle and right panels of Figure \ref{fig:NSB sigma 0} show the post-kick distributions of $a_\MID^k$ and $e_\MID^k$.
We see that the systems that survived the kicks can cover near the whole parameter space in the
middle panel.
If the natal kick for the newly formed BH ($m_3$) is increased to $\sigma_\mathrm{BH}=~100\mathrm{km~s^{-1}}$,
a significant reduction of survivals can be found, similar to the case depicted in Figure \ref{fig:BHB sigma 0 100}.

To determine the merger fractions of the middle binaries,
we carry out calculations similar to the one in Section \ref{sec 3 3}.
The results are shown in Figure \ref{fig:Merger fraction in middle NSB}.
We find that compared to the example in Figure \ref{fig:Merger fraction in middle BHB},
the parameter region with high merger fraction ($f_\merger^\m\gtrsim10\%$, $5\%$) is narrower.
This is because the component masses of the middle binary ($m_{12}=2.6M_\odot$ and $m_3=23.2M_\odot$) are smaller, GW radiation becomes less efficient,
reducing the number of systems which are able to merge within a Hubble time.

\begin{figure}
\centering
\begin{tabular}{cccc}
\includegraphics[width=8cm]{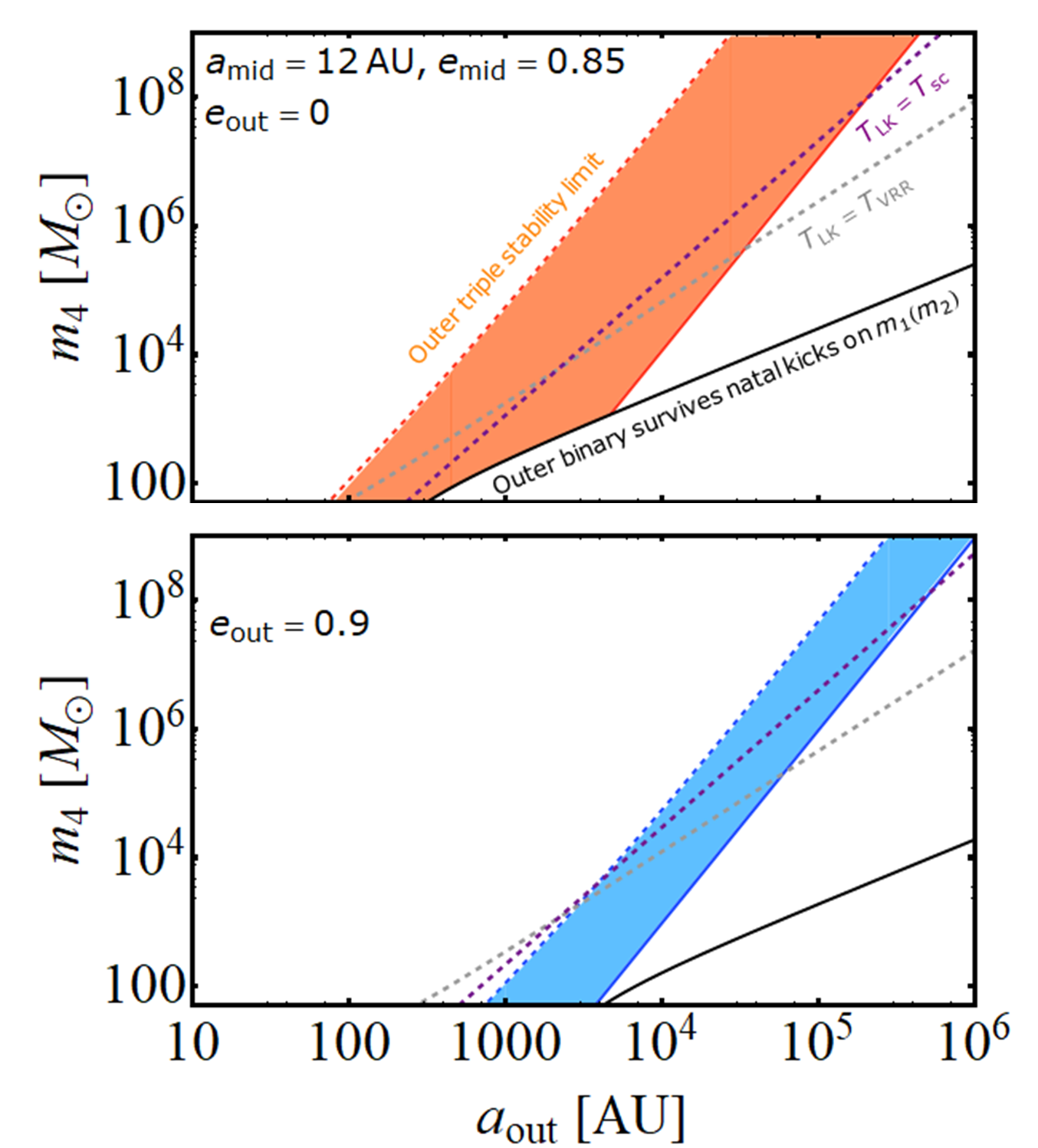}
\end{tabular}
\caption{Similar to Figure \ref{fig:m4 orbit}, but for GW190812-like events. The representative middle binary has $m_{12}=2.6M_\odot$, $m_3=23.2M_\odot$,
$a_\MID=12\au$ and $e_\MID=0.85$.
}
\label{fig:m4 orbit NSB}
\end{figure}

In Figure \ref{fig:m4 orbit NSB}, we show the constrained parameter space for the tertiary companion $m_4$ (MBH), where
we consider the representative middle binary orbital parameters
$a_\MID=12\au$ and $e_\MID=0.85$ (see the star symbol shown in Figure \ref{fig:Merger fraction in middle NSB}).
Different from the case of GW190412,
the middle binary here is affected by the large natal kicks on $m_1$ or $m_2$ (instead of the merger kick on $m_{12}$),
and varying the middle binary property can influence the constraint on the outer binary.
Nevertheless, we see from Figure \ref{fig:m4 orbit NSB} that in general,
a large tertiary mass (from $100M_\odot$ to $\gtrsim10^8M_\odot$) is required to induce merger of the GW190814-like BHBs.

\section{Formation of GW190521}
\label{sec 5}

Since the two BHs in GW190521 likely fall in the high mass gap, and each BH has a dimensionless spin parameter close to $\sim0.7$,
we suggest that both the components in this source are 2G BHs and are produced by previous equal-mass mergers (see Figure \ref{fig:m1m2}).
Therefore, in our ``primordial" multiple scenario (see the left pathway in Figure \ref{fig:Formation of middle BHB GW190521}),
we consider a quadruple system consisting of two inner binaries, and the progenitor masses are
($47.2 M_\odot$, $47.2 M_\odot$) and ($36.7 M_\odot$, $36.7 M_\odot$).
We assume that the system goes through SN explosion on each component with a negligible natal kick
($\sigma_\mathrm{BH}=0~\mathrm{km~s^{-1}}$) and sudden mass loss, as well as two merger kicks with
$V_\mathrm{mk}\simeq0~\mathrm{km~s^{-1}}$ (because of the equal-mass merger; see Figure \ref{fig:m1m2}). Eventually,
the two remnants form the middle binary, which then merge with the aid of an external companion
$m_5$ (the MBH; see Figure \ref{fig:Formation of middle BHB GW190521}).

For the survival fraction of the middle binaries, different from GW190412 and GW190814, all kicks are negligible in the case of GW19052.
Thus, we expect that almost all the middle binaries satisfying the stability criterion can
survive with high probabilities.
The distribution of the post-kick $a_\MID^k$ and $e_\MID^k$ is expected to be similar to Figure \ref{fig:BHB sigma 0 case II}
(since Figure \ref{fig:BHB sigma 0 case II} also corresponds to the case with small natal kick and merger kick).

\begin{figure}
\centering
\begin{tabular}{cccc}
\includegraphics[width=8.5cm]{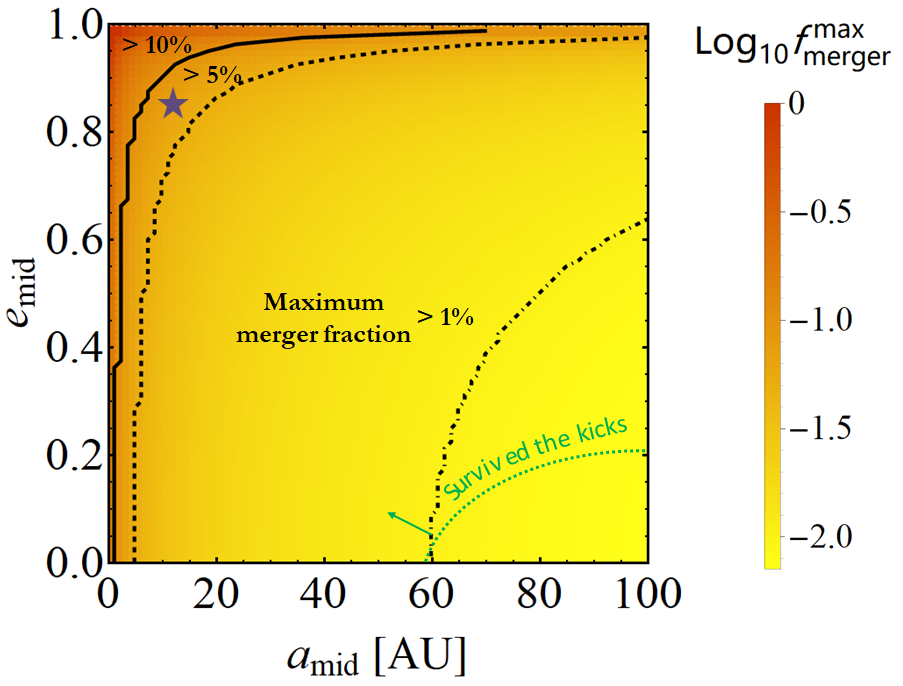}
\end{tabular}
\caption{Similar to Figure \ref{fig:Merger fraction in middle BHB}, but for the
middle binary with $m_{12}=2.6M_\odot$, and $m_3=23.2M_\odot$
(appropriate for GW190814-like events). The green dashed line is the same as the one in the middle panel of Figure \ref{fig:NSB sigma 0}.
}
\label{fig:Merger fraction in middle NSB}
\end{figure}

\begin{figure}
\centering
\begin{tabular}{cccc}
\includegraphics[width=8.5cm]{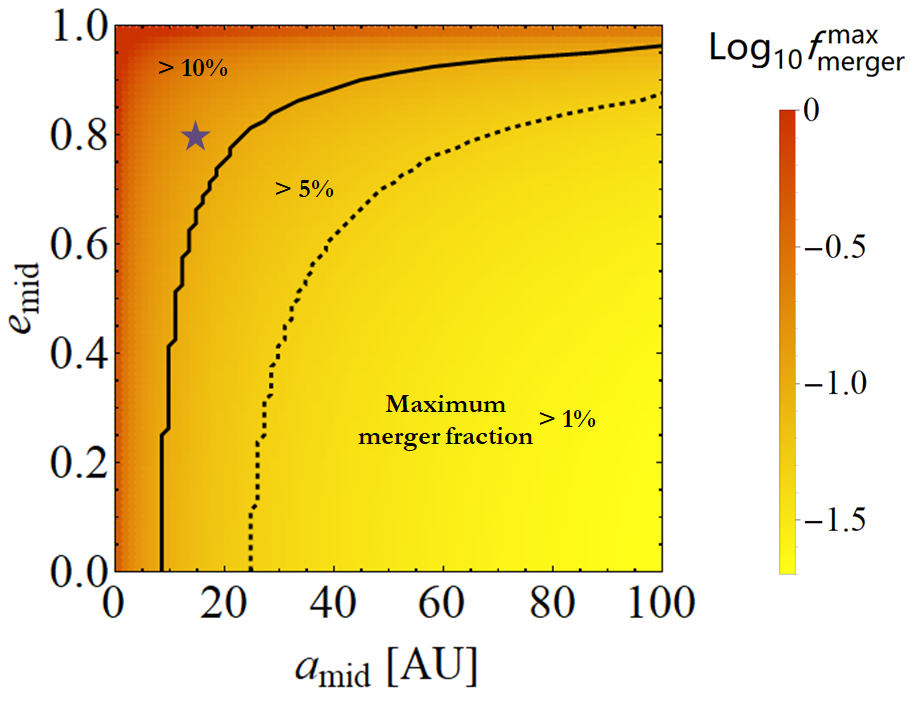}
\end{tabular}
\caption{Similar to Figure \ref{fig:Merger fraction in middle BHB}, but for the
middle binary with $m_{12}=85M_\odot$, and $m_3=66M_\odot$
(appropriate for GW190521-like events). Since both the natal kick and merger kick are negligible, all ($a_\MID$, $e_\MID$) values can survive the kicks.
}
\label{fig:Merger fraction in middle BHB GW190521}
\end{figure}

For the merger fractions, we use the same approach as in Section \ref{sec 3 3} and
present the results in Figure \ref{fig:Merger fraction in middle BHB GW190521}.
Since the component masses of the middle binary are large ($85M_\odot$ and $66 M_\odot$),
we find that the parameter region with $f_\merger^\m\gtrsim10\%$ ($5\%$) are the broader compared to
the other examples in Figures \ref{fig:Merger fraction in middle BHB} and \ref{fig:Merger fraction in middle NSB}.

\begin{figure}
\centering
\begin{tabular}{cccc}
\includegraphics[width=8cm]{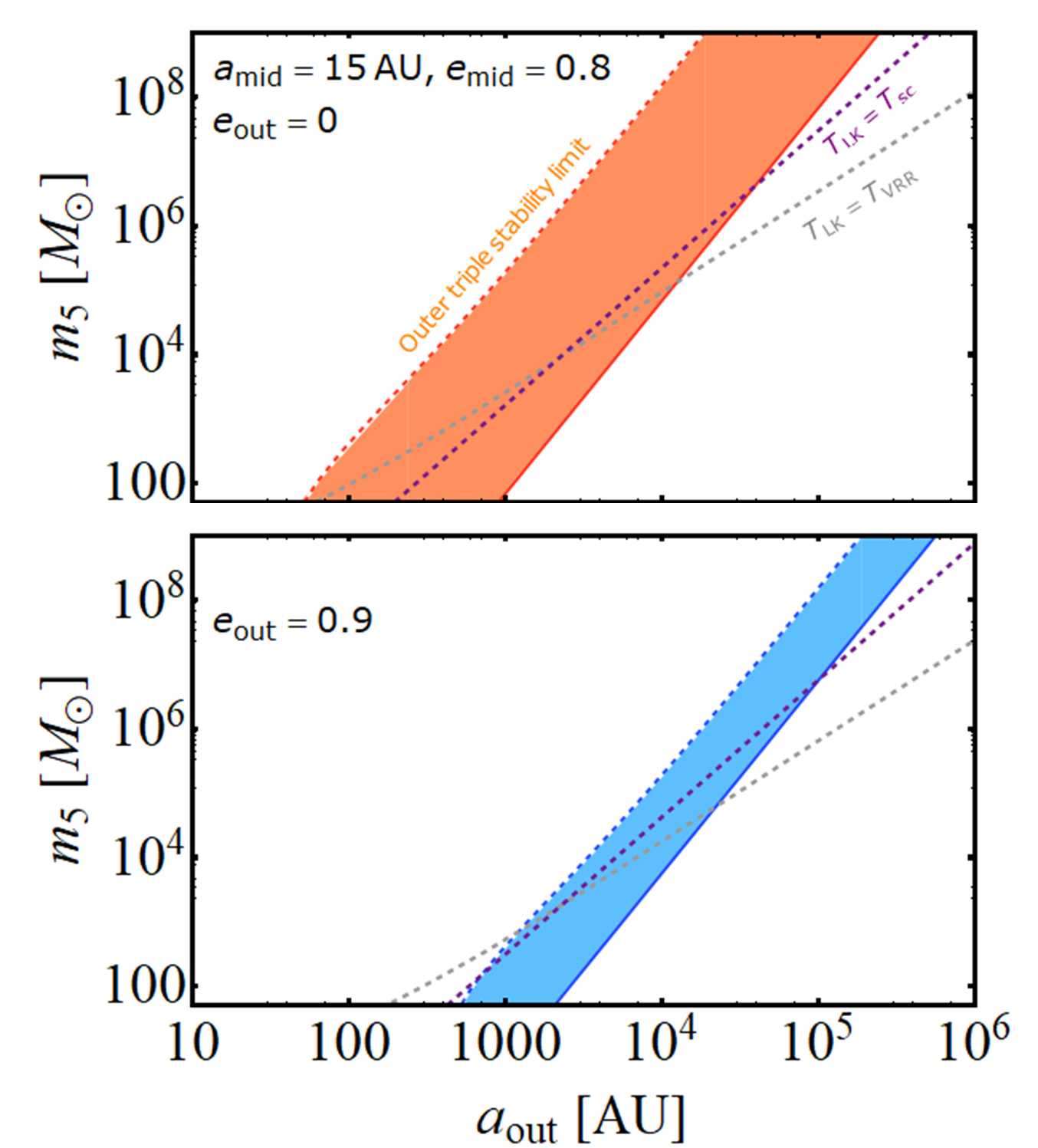}
\end{tabular}
\caption{Similar to Figure \ref{fig:m4 orbit}, but for GW190521-like events.
The representative middle binary has $m_{12}=85M_\odot$, $m_3=66M_\odot$,
$a_\MID=15\au$ and $e_\MID=0.8$. Note that since the kicks are negligible, there are no constraints from kicks on the tertiary companion.
}
\label{fig:m4 orbit GW190521}
\end{figure}

We show the constrained parameters for the external companion (i.e., the fifth body $m_5$, the MBH) in
Figure \ref{fig:m4 orbit GW190521}.
Here, we consider the representative  middle binary orbital parameters
$a_\MID=15\au$ and $e_\MID=0.8$ (see the star symbol shown in Figure \ref{fig:Merger fraction in middle BHB GW190521}).
Note that in this case, there are no constraints on $m_5$ and $a_\OUT$ from the natal kick and merger kick.
Again, we find that a broad range of $m_5$ values are possible.

\section{Dynamically Formed BH Binaries}
\label{sec 6}

In Sections \ref{sec 3}-\ref{sec 5} we have studied in detail the
formation of GW190412, GW190814 and GW190521-like systems assuming that the BHBs
(the middle binary in Figure \ref{fig:configuration}) are produced in ``primordial" multiple systems
(the leftmost pathway in Figures \ref{fig:Formation of middle BHB GW190412}-\ref{fig:Formation of middle BHB GW190521}).
As noted in Section \ref{sec 1}, these BHBs can also form dynamically, especially when the external companion is a SMBH.
Indeed, the kinds of systems we study may be naturally found in the nuclear star cluster around a SMBH.
Recent studies show that hierarchical systems of stars
and compact remnants can form through few-body encounters in dense star clusters,
either through binary-single interaction \citep[e.g.,][]{Samsing 2014} or through binary-binary interaction \citep[e.g.,][]{Fragione 2020}.

Suppose the BHBs leading to the three LIGO/VIRGO events could be assembled dynamically in nuclear star clusters,
the masses of their progenitor systems are constrained by two factors:
(i) In typical binary-single (or binary) encounters, a single (or a tight binary) is exchanged into a bound orbit,
while the lightest mass is ejected;
(ii) If a binary component is a merger remnant and its mass and spin are known from the GW data
(as in the case of GW190412), then the masses of its progenitors can be constrained (see Section \ref{sec 3 1}).

Figures \ref{fig:Formation of middle BHB GW190412}-\ref{fig:Formation of middle BHB GW190521} show the possible dynamical pathways
leading to the formation of BHBs in the three LIGO/VIRGO O3 events.
Taking the GW190412 as an example (Figure \ref{fig:Formation of middle BHB GW190412}),
a compact BHB with mass ($25.2 M_\odot$, $5.3 M_\odot$) or ($21.9 M_\odot$, $9.2 M_\odot$)
(see Case I and III in Table \ref{tab:Parameters of Progenitors}) merges first.
Then, the newly form (2G) BH interacts with a binary ($8 M_\odot$, $4 M_\odot$),
leaving behind a BHB with mass ($30 M_\odot$, $8 M_\odot$) and an escaper $4 M_\odot$.
Alternatively, the compact ($25.2 M_\odot$, $5.3 M_\odot$) or ($21.9 M_\odot$, $9.2 M_\odot$) binary
may experience a close encounter with the wide ($8 M_\odot$, $4 M_\odot$) binary first,
replacing the light component in the wide binary.
Then, the compact binary in this triple can merge (either by itself, or through LK oscillations), leading to the 2G BH ($30 M_\odot$).
Eventually the GW190412-like ``middle" binary is formed.

Numerous uncertainties exist in these dynamical scenarios, including the formation fraction and survival fraction of the BHBs (the middle binaries).
The former depends on the density, initial mass function and binary abundance in the cluster.
The latter is mostly determined by the merger kick (received by the 2G BH), since
a large fraction of the dynamically formed triple/binary population consists of BHs
\citep[e.g.,][]{Fragione 2020}. These uncertainties imply that the distributions of the survived middle binaries
in $a_\MID^k-e_\MID^k$ parameter plane are highly uncertain.

Nevertheless, we expect that a broad range of $a_\MID$ and $e_\MID$ can be produced dynamically.
The merger fractions shown in Figures \ref{fig:Merger fraction in middle BHB}, \ref{fig:Merger fraction in middle NSB} and
\ref{fig:Merger fraction in middle BHB GW190521} still apply
to the dynamically formed BHBs, except that the green ``survived kicks" lines in Figures \ref{fig:Merger fraction in middle BHB} and
\ref{fig:Merger fraction in middle NSB} are irrelevant.
Similarly, Figures \ref{fig:m4 orbit}, \ref{fig:m4 orbit NSB}, \ref{fig:m4 orbit GW190521} are still valid for dynamically formed BHBs,
except that the ``outer binary survives merger (natal) kick" lines in Figures \ref{fig:m4 orbit} and \ref{fig:m4 orbit NSB} are irrelevant.
Overall, these figures provide constraints on the masses of the MBH and the locations where tertiary-induced mergers
can happen leading to these three LIGO/VIRGO events.

\section{Discussion}
\label{sec 7}

\subsection{How common are GW190412, GW190814 and GW190521-like events?}
\label{sec 7 1}

The main goal of this paper is to constrain what kinds of binary/triple/quadrupole systems (e.g. the range of $a_\MID$ and $m_\mathrm{MBH}$)
can lead to GW190412, GW190814 and GW190521-like events in the hierarchical tertiary-induced merger scenario.
Given the uncertainties in the binary/triple/quadrupole populations (especially those in dense star clusters), a systematical evaluation of the
occurrence rates of such events is beyond the scope of this paper --- our calculations presented in previous sections were not
set up for population synthesis study. Indeed, our view is that full population synthesis study
(starting from main sequence stellar binaries and triples) would involve too many uncertainties to be valuable at this point.
Nevertheless, it is of interest to have a crude, back-of-the-envelope estimates of the merger rates for
GW190412, GW190814 and GW190521-like events in our scenario. We carry out such estimate in the following.
Since we do not consider full stellar evolution (except SNe and natal kicks) in our study,
we evaluate the merger rate of these 2G BHs using the merger rate ($\mathcal{R}_\mathrm{1G}$) of 1G BHs
(or NSs in the case GW190814),
which we assume to be a fraction of the detection rate of ``normal" BH or NS mergers from LIGO/VIRGO O1-O2 runs.
Although our analysis in Sections \ref{sec 3}-\ref{sec 5} are quite general,
the results indicate that the tertiary companion is likely a MBH.
We therefore envision that these events likely take place in dense star clusters around MBHs.

A key uncertainty in our rate estimate is the ``primordial" stellar or BH multiplicity.
It is shown observationally that the majority of massive stars
are born in binaries or higher-order multiple systems \citep[e.g.,][]{Sana 2012}.
The situation in nuclear star clusters is unclear, but we expect that even higher multiple fractions are possible. We assume that the
formation fraction of stellar triples $f_\mathrm{triple}$ is about $50\%$.

First consider the GW190412-like events (see Figure \ref{fig:Formation of middle BHB GW190412} or \ref{fig:configuration}).
The BHBs (middle binaries) in our simulations can have significant survival fraction and merger fraction, i.e.,
$f_\mathrm{survival}$, $f_\merger\gtrsim10\%$, when the BH natal kicks are negligible (see Sections \ref{sec 3 2} and \ref{sec 3 3}).
Therefore, the merger rate of 2G BHs can be estimated as
\footnote{
Note that in Equation (\ref{eq:merger rate BHB}), we have adopted $f_\mathrm{survival} \sim 10\%$
based on our canonical calculation in Section \ref{sec 3 2}
(see Figure \ref{fig:BHB sigma 0 case I}).
If the triple population contains systems with $a_\MID^0$ beyond 100 AU, $f_\mathrm{survival}$ would be reduced by a factor of $\sim 2$
\citep[note that the $a_\MID^0$ distribution of stellar binaries beyond 100 AU is steeper than $(a_\MID^0)^{-1.5}$; see][]{Moe 2017,El 2018,Tian 2020}.
On the other hand, if the triple population in dense star cluster favors systems with $a_\MID^0$ much smaller than 100 AU, $f_\mathrm{survival}$
could be increased somewhat. In any case, what matter is the product $f_\mathrm{triple}\times f_\mathrm{survival}$, which we have taken to be
$5\%$ in Equation (\ref{eq:merger rate BHB}).
}:
\be\label{eq:merger rate BHB}
\begin{split}
\mathcal{R}_\mathrm{2G}&=\mathcal{R}_\mathrm{1G}\times f_\mathrm{triple}\times f_\mathrm{survival}\times f_\mathrm{merger}\\
&\sim \mathcal{R}_\mathrm{1G}\times50\%\times10\%\times10\%\\
&\simeq 0.5\%\times \mathcal{R}_\mathrm{1G}.
\end{split}
\ee
If we take the merger rate of 1G BHs in our scenario, $\mathcal{R}_\mathrm{1G}$, to be fraction of the LIGO O1-O2 detection rate,
$\mathcal{R}_\mathrm{obs}=10-100 \mathrm{Gpc}^{-3}\mathrm{yr}^{-1}$ \citep[][]{LIGO 2019a},
we find $\mathcal{R}_\mathrm{2G}\lesssim0.05-0.5\mathrm{Gpc}^{-3}\mathrm{yr}^{-1}$.
Note that $\mathcal{R}_\mathrm{2G}$ will be lower if the natal kicks on the BHs are significant,
as the binary survival fraction will be decreased.
On the other hand, various ``environmental" effects may increase the merger fraction $f_\mathrm{merger}$ by a factor of a few
(see Section \ref{sec 7 2}).
For reference,
\citet{GW190412} suggested that the merger rate of BHBs with mass ratio $m_2/m_1\lesssim0.4$ is about $10\%$ of the O1-O2 rate,
i.e., $1-10 \mathrm{Gpc}^{-3}\mathrm{yr}^{-1}$, and the rate for BHB mergers with $m_2/m_1\lesssim0.25$ (like GW190412) is even lower
\citep[see also][]{Olejak 2020}.
Our upper limit for $\mathcal{R}_\mathrm{2G}$ for GW190412-like events
is nominally smaller than the ``observed" LIGO rate, but given the various uncertainties, it could also be consistent with the LIGO
rate.

For the GW190814-like events (see Figure \ref{fig:Formation of middle BHB GW190814}),
we again set $f_\mathrm{survival}$, $f_\merger\simeq10\%$
and obtain $\mathcal{R}_\mathrm{2G}\sim 0.5\% \times \mathcal{R}_\mathrm{1G}$.
The observed rate of 1G NS binaries suggested by LIGO is about $\mathcal{R}_\mathrm{obs}=250-2810~\mathrm{Gpc}^{-3}\mathrm{yr}^{-1}$
\citep[e.g.,][]{GW190425},
so our inferred merger rate induced by a tertiary is $\mathcal{R}_\mathrm{2G}\lesssim1-14 \mathrm{Gpc}^{-3}\mathrm{yr}^{-1}$.
We emphasize that this rate estimate is highly uncertain since the survival fraction
depends sensitively on the initial configuration of the system, where the large natal kicks on newly born NSs can destroy the ``middle" binaries.
For reference, the LIGO detection rate of GW190814-like events
is about $\sim1-23 \mathrm{Gpc}^{-3}\mathrm{yr}^{-1}$ \citep[][]{GW190814}, and our estimate for
$\mathcal{R}_\mathrm{2G}$ is consistent with this ``observed" value.

For the GW190521-like binaries (see Figure \ref{fig:Formation of middle BHB GW190521}),
the survival fraction and merger fraction can be higher compared to the other sources
if we adopt negligible natal kick and merger kick (see Section \ref{sec 5}).
Assuming an upper limit $f_\mathrm{survival}\simeq60\%$ and $f_\merger\simeq20\%$ (based on the discussion in Section \ref{sec 5}),
we find $\mathcal{R}_\mathrm{2G} \sim 6\% \times \mathcal{R}_\mathrm{1G}$.
Using $\mathcal{R}_\mathrm{1G}\lesssim10-100 \mathrm{Gpc}^{-3}\mathrm{yr}^{-1}$, we find
$\mathcal{R}_\mathrm{2G}\lesssim0.6-6~\mathrm{Gpc}^{-3}\mathrm{yr}^{-1}$.
Note that this is an optimistic estimate: The survival fraction can be lower if the binary components come
from the mergers of BHB with highly asymmetric masses (i.e., merger kick can be large).
For reference, the merger rate given by LIGO/VIRGO for GW190521-like events
is $\sim0.02-0.43 \mathrm{Gpc}^{-3}\mathrm{yr}^{-1}$ \citep[][]{GW190521}.

Overall, although the above estimates indicate
that the rate for tertiary-induced mergers in multiples could be consistent with the LIGO/VIRGO findings, our estimates could be too optimistic.
The merger rate of 1G BHs near the MBH/SMBH is quite uncertain.
Also, the survival fraction can be lower if the natal kick of newly born BH is large,
leading to smaller merger rate estimates.

\subsection{``Environmental" Effects}
\label{sec 7 2}

Our analyses in Sections \ref{sec 3}-\ref{sec 5}
do not take into account the effects related to dense stellar environment.
Since we have found that the external companion in our scenario is likely a MBH,
the effects from the cluster may play an important role and change the configurations of the inner triple systems studied here.
We list three main effects as follows.

In a dense stellar environment, the BHB (the middle binary) may be perturbed through multiple encounters with
other passing objects. The orbital parameters may change significantly, leading to binary evaporation.
The typical timescale can be estimated as \citep[e.g.,][]{Binney and Tremaine}
\be
T_\mathrm{evap}=\frac{(m_{12}+m_3)\sigma_\mathrm{cl}}{16\sqrt{\pi}Ga_\MID\langle m\rangle\rho_\mathrm{cl}\mathrm{ln}\Lambda}.
\ee

To evaluate $T_\mathrm{evap}$,
we assume the stellar velocity dispersion $\sigma_\mathrm{cl}=280\mathrm{km~s^{-1}}\sqrt{0.1\mathrm{pc}/a_\OUT}$
\citep[e.g.,][]{Kocsis 2011},
the stellar density $\rho_\mathrm{cl}=0.8\times10^5m_\odot\mathrm{pc}^{-3}(a_\OUT/\mathrm{pc})^{-1.3}$
\citep[approproate for Mikly Way galactic center;][]{Fritz 2016},
the averaged stellar mass in the cluster $\langle m\rangle=1M_\odot$, and the Coulomb logarithm $\mathrm{ln}\Lambda\simeq2$.
In the examples of Figures \ref{fig:m4 orbit}, \ref{fig:m4 orbit NSB} and \ref{fig:m4 orbit GW190521}, we find that $T_\mathrm{evap}$ is
always greater than the LK timescale $T_\lk$ if $a_\OUT\lesssim10^6\au$
\footnote
{Here, we only consider one example of the galactic center, and do not take into account the possible
dependence of $\sigma_\mathrm{cl}$ and $\rho_\mathrm{cl}$ on $m_4$ (that is complicated and highly uncertain).
In some of the nuclear star cluster, $T_\mathrm{evap}$ could be short and BHBs may not survive for a sufficiently long time.
}.
Therefore, the MBH can induce LK merger of the middle binary
before the disruption of middle binary (color-shaded region).

Another effect is resonant relaxation, which affects orbits close to the MBH,
and changes both the magnitude and direction of the orbital angular momentum \citep[e.g.,][]{Rauch 1996}.
The timescale for the relaxation of the orbital orientation vectors
(the ``Vector Resonant Relaxation", or VRR) is given by \citep[e.g.,][]{Hamers 2018}
\be\label{eq:VRR timescale}
T_\mathrm{VRR}=\frac{P_\OUT}{\beta}\frac{m_\mathrm{MBH}}{\sqrt{N_\mathrm{cl}}},
\ee
where $\beta\sim 2$, $P_\OUT$ is the orbital period of the outer binary, $N_\mathrm{cl}$
is the number of stars within radius $r=a_\OUT$ from the MBH.
Because VRR can change the orientation of BHB$+$SMBH binary orbit,
it can enhance the tertiary-induced merger rate by opening up the LK window through an ``inclination resonance"
\citep[e.g.,][]{Hamers and Lai 2017,Hamers 2018}, provided that $T_\mathrm{VRR}$ is less than $T_\lk$.
The grey dashed lines in Figures \ref{fig:m4 orbit}, \ref{fig:m4 orbit NSB} and \ref{fig:m4 orbit GW190521} indicate
$T_\mathrm{VRR}=T_\lk$, highlighting the systems
that might have such ``inclination resonance".

Finally, if the nuclear star cluster has a non-spherical mass distributions, the BHB's orbit around the central MBH
will experience nodal precession induced by the non-spherical cluster potential.
The characteristic timescale is \citep[e.g.,][]{Petrovich 2017}
\be\label{eq:cluster potential}
T_\mathrm{sc}=\frac{1}{\epsilon_z GP_\OUT\rho_\mathrm{cl}},
\ee
where $\epsilon_z$ is dimensionless and measures the asphericity of the cluster mass distribution.
Thus, for a BHB embedded in such environment, the angular momentum of the orbit around the MBH may
vary in time, changing the inclination angle between the inner and outer orbits.
When the precession time $T_\mathrm{sc}$ is comparable to the LK timescale $T_\lk$,
the merger window of the BHB induced by the SMBH can increase as a result of the ``inclination resonance",
further enhancing the BHB merger rate \citep[][]{Petrovich 2017}.
The ``$T_\mathrm{sc}=T_\lk$" lines (assuming $\epsilon_z\sim 0.1$) in
Figures \ref{fig:m4 orbit}, \ref{fig:m4 orbit NSB} and \ref{fig:m4 orbit GW190521} indicate the systems for which such enhancements are effective.

Overall, these dynamical effects of the cluster environment can increase the merger fraction of BHBs induced by the SMBH
from $10\%$ to more than $30-50\%$, thus increasing our rate estimates in Section \ref{sec 7 1} by a factor of a few.

\section{Summary}
\label{sec 8}

We have studied the formation of three exceptional LIGO/VIRGO O3
events, GW190412, GW190814 and GW190521, in hierarchical multiple
systems, where one or both components of the binary come from a
previous merger. In our scenario, the multiples could be either
``primordial'' or formed dynamically in dense stellar clusters (see
Figures \ref{fig:Formation of middle BHB GW190412}-\ref{fig:Formation of middle BHB GW190521}).
Regardless of the detailed evolutionary pathways, the
final black-hole binaries (BHBs) generally have too wide orbital
separations to merge by themselves. Instead, with the aid of an
external companion (likely a massive black hole, MBH), the binary can
merge over the cosmic time due to Lidov-Kozai (LK) eccentricity
oscillations. During its evolution, the progenitor multiple system
undergoes supernova explosions to form BHs or NSs accompanied by mass
losses and natal kicks, and may also experience kick when the
first-generation BHs merge. All these kicks and mass losses can
change the configuration/geometry of the hierarchical multiple, or even
break up the system. We explore the binaries that are most likely to
survive the kicks, and use the post-kick binary distributions to
constrain the parameter space of the external companion which can give
rise to LK-induced BHB mergers.

The results of our calculations can be summarized as follows.

(i) For the inner triple systems in our ``primordial'' multiple
scenario (Figure \ref{fig:configuration}; see also the leftmost pathway in
Figures \ref{fig:Formation of middle BHB GW190412}-\ref{fig:Formation of middle BHB GW190521}), only the relative compact
middle binaries can survive both natal and merger kicks, even though
the natal kick on a newly born BH may be negligible (see the left panels of
Figures \ref{fig:BHB sigma 0 case I} and \ref{fig:NSB sigma 0}).
The distributions of the post-kick orbital parameters of all survivals
show that the majority of systems have either small semimajor axes or
large eccentricities (see the middle and right panels of Figures
\ref{fig:BHB sigma 0 case I} and \ref{fig:NSB sigma 0}).  This is
partly because the wide binaries can be easily destroyed and partly
because the sudden mass loss can increase the orbital eccentricity.

(ii) For the LK-induced BHB mergers with sufficiently massive tertiary
(such as MBH), the octupole effects are negligible, and the merger fraction
(assuming the orientation of the tertiary is isotropically distributed)
as a function of the binary and tertiary parameters can be determined analytically
(Section \ref{sec 2 2}, Figure \ref{fig:merger window}). The maximum merger fraction ($f_\merger^\m$) is given by
the analytical formula (\ref{eq:fitting f merger}).
The value of $f_\merger^\m$ depends only on the properties of the BHB,
instead of the tertiary companion.
The strength of the tertiary perturbation can be characterized by the dimensionless effective
semi-major axis $\bar{a}_{\OUT,\eff}$ (see Equation \ref{eq:aout bar}).
The weakest perturbation (the largest $\bar{a}_{\OUT,\eff}$),
beyond which no merger can occur, is given by the analytical formula (\ref{eq:fitting a out}).

(iii) Based on the systems that have survived SNe and kicks,
together with our analytical formulae for LK-induced mergers, we constrain the
the properties of the tertiary companions required to produce the three LIGO/VIRGO O3
events (see Figures \ref{fig:m4 orbit}, \ref{fig:m4 orbit NSB} and \ref{fig:m4 orbit GW190521}).
These constraints indicate that the tertiary companions must be at least
a few hundreds $M_\odot$, and fall in the intermediate-mass BH and supermassive BH range.

(iv) Based on our calculations, we suggest that GW190412, GW190814 and
GW190521 could all be produced via hierarchical mergers in multiples:
Through different evolutionary pathways
(Figures \ref{fig:Formation of middle BHB GW190412}-\ref{fig:Formation of middle BHB GW190521}),
a BHB is assembled, likely
in a dense nuclear star cluster, and the final merger is induced by a
MBH or SMBH. We estimate the event rates of such hierarchical mergers
based on the ``normal'' BHB or NS binary merger rates from LIGO/VIRGO
O1-O2 runs and our calculated binary survival and merger fractions.
The merger rate given by this tertiary-induced channel is very uncertain. But our optimistic estimate indicates
that the rate could be consistent with the LIGO/VIRGO findings (Section \ref{sec 7 1}).

\section{Acknowledgments}

This work is supported in part by the NSF grant AST-1715246.

\section{DATA AVAILABILITY}

The simulation data underlying this article will be shared
on reasonable request to the corresponding author.

\label{lastpage}


\begin{thebibliography}{}

\bibitem[\protect\citeauthoryear{Abbott et al.}{2019a}]{LIGO 2019a}
Abbott B.~P., Abbott R., Abbott T.~D., Abraham S., Acernese F., Ackley K., Adams C., et al., 2019, PhRvX, 9, 031040

\bibitem[\protect\citeauthoryear{Abbott et al.}{2019b}]{LIGO 2019b}
Abbott B.~P., Abbott R., Abbott T.~D., Abraham S., Acernese F., Ackley K., Adams C., et al., 2019, ApJL, 882, L24

\bibitem[\protect\citeauthoryear{Abbott et al.}{2020a}]{GW190412}
Abbott R., Abbott T.~D., Abraham S., Acernese F., Ackley K., Adams C., Adhikari R.~X., et al., 2020, PhRvD, 102, 043015

\bibitem[\protect\citeauthoryear{Abbott et al.}{2020b}]{GW190425}
Abbott B.~P., Abbott R., Abbott T.~D., Abraham S., Acernese F., Ackley K., Adams C., et al., 2020, ApJL, 892, L3

\bibitem[\protect\citeauthoryear{Abbott et al.}{2020c}]{GW190814}
Abbott R., Abbott T.~D., Abraham S., Acernese F., Ackley K., Adams C., Adhikari R.~X., et al., 2020, ApJL, 896, L44

\bibitem[\protect\citeauthoryear{Abbott et al.}{2020d}]{GW190521}
Abbott R., Abbott T.~D., Abraham S., Acernese F., Ackley K., Adams C., Adhikari R.~X., et al., 2020, PhRvL, 125, 101102

\bibitem[\protect\citeauthoryear{Abbott et al.}{2020e}]{GW190521 ApJL}
Abbott R., Abbott T.~D., Abraham S., Acernese F., Ackley K., Adams C., Adhikari R.~X., et al., 2020, ApJL, 900, L13

\bibitem[\protect\citeauthoryear{Anderson et al.}{2016}]{Anderson et al 2016}
Anderson K.~R., Storch N.~I., Lai D., 2016, MNRAS, 456, 3671

\bibitem[\protect\citeauthoryear{Anderson et al.}{2017}]{Anderson et al 2017}
Anderson K.~R., Lai D., Storch N.~I., 2017, MNRAS, 467, 3066

\bibitem[\protect\citeauthoryear{Antonini \& Perets}{2012}]{Antonini 2012}
Antonini F., Perets H.~B., 2012, ApJ, 757, 27

\bibitem[\protect\citeauthoryear{Antonini et al.}{2017}]{Antonini 2017}
Antonini F., Toonen S., Hamers A.~S., 2017, ApJ, 841, 77

\bibitem[\protect\citeauthoryear{Bailyn et al.}{1998}]{Bailyn 1998}
Bailyn C.~D., Jain R.~K., Coppi P., Orosz J.~A., 1998, ApJ, 499, 367

\bibitem[\protect\citeauthoryear{Banerjee et al.}{2010}]{Banerjee 2010}
Banerjee S., Baumgardt H., Kroupa P., 2010, MNRAS, 402, 371

\bibitem[\protect\citeauthoryear{Barkat et al.}{1967}]{Barkat 1967}
Barkat Z., Rakavy G., Sack N., 1967, PhRvL, 18, 379

\bibitem[\protect\citeauthoryear{Bartos et al.}{2017}]{Bartos}
Bartos I., Kocsis B., Haiman Z., M{\'a}rka S., 2017, ApJ, 835, 165

\bibitem[\protect\citeauthoryear{Belczynski et al.}{2010}]{Belczynski 2010}
Belczynski K., Dominik M., Bulik T., O'Shaughnessy R., Fryer C., Holz D.~E., 2010, ApJ, 715, L138

\bibitem[\protect\citeauthoryear{Belczynski et al.}{2016}]{Belczynski 2016}
Belczynski K., Holz D.~E., Bulik T., O'Shaughnessy R., 2016, Natur, 534, 512

\bibitem[\protect\citeauthoryear{Belczynski et al.}{2018}]{Belczynski 2018}
Belczynski K., Askar A., Arca-Sedda M., Chruslinska M., Donnari M., Giersz M., Benacquista M., et al., 2018, A\&A, 615, A91

\bibitem[\protect\citeauthoryear{Binney \& Tremaine}{1987}]{Binney and Tremaine}
Binney J., Tremaine S., 1987, gady.book

\bibitem[\protect\citeauthoryear{Blaes et al.}{2002}]{Blaes 2002}
Blaes O., Lee M.~H., Socrates A., 2002, ApJ, 578, 775

\bibitem[\protect\citeauthoryear{Gerosa et al.}{2020}]{Gerosa 2020}
Gerosa D., Vitale S., Berti E., 2020, PhRvL, 125, 101103

\bibitem[\protect\citeauthoryear{Di Carlo et al.}{2020}]{Di Carlo 2020}
Di Carlo U.~N., Mapelli M., Giacobbo N., Spera M., Bouffanais Y., Rastello S., Santoliquido F., et al., 2020, arXiv, arXiv:2004.09525

\bibitem[\protect\citeauthoryear{Downing et al.}{2010}]{Downing 2010}
Downing J.~M.~B., Benacquista M.~J., Giersz M., Spurzem R., 2010, MNRAS, 407, 1946

\bibitem[\protect\citeauthoryear{Dominik et al.}{2012}]{Dominik 2012}
Dominik M., Belczynski K., Fryer C., Holz D.~E., Berti E., Bulik T., Mandel I., O'Shaughnessy R., 2012, ApJ, 759, 52

\bibitem[\protect\citeauthoryear{Dominik et al.}{2013}]{Dominik 2013}
Dominik M., Belczynski K., Fryer C., Holz D.~E., Berti E., Bulik T., Mandel I., O'Shaughnessy R., 2013, ApJ, 779, 72

\bibitem[\protect\citeauthoryear{Dominik et al.}{2015}]{Dominik 2015}
Dominik M., et al., 2015, ApJ, 806, 263

\bibitem[\protect\citeauthoryear{El-Badry \& Rix}{2018}]{El 2018}
El-Badry K., Rix H.-W., 2018, MNRAS, 480, 4884

\bibitem[\protect\citeauthoryear{Fabrycky \& Tremaine}{2007}]{Fabrycky 2007}
Fabrycky D., Tremaine S., 2007, ApJ, 669, 1298

\bibitem[\protect\citeauthoryear{Farr et al.}{2011}]{Farr 2011}
Farr W.~M., Sravan N., Cantrell A., Kreidberg L., Bailyn C.~D., Mandel I., Kalogera V., 2011, ApJ, 741, 103

\bibitem[\protect\citeauthoryear{Ford et al.}{2000}]{Ford}
Ford E. B., Kozinsky B., Rasio F. A., 2000b, ApJ, 535, 385

\bibitem[\protect\citeauthoryear{Fragione \& Kocsis}{2019a}]{Fragione 2019a}
Fragione G., Kocsis B., 2019, MNRAS, 486, 4781

\bibitem[\protect\citeauthoryear{Fragione \& Loeb}{2019b}]{Fragione 2019b}
Fragione G., Loeb A., 2019, MNRAS, 486, 4443

\bibitem[\protect\citeauthoryear{Fragione \& Loeb}{2019c}]{Fragione 2019c}
Fragione G., Loeb A., 2019, MNRAS, 490, 4991

\bibitem[\protect\citeauthoryear{Fragione et al.}{2020}]{Fragione 2020}
Fragione G., Martinez M.~A.~S., Kremer K., Chatterjee S., Rodriguez C.~L., Ye C.~S., Weatherford N.~C., et al., 2020, arXiv:2007.11605

\bibitem[\protect\citeauthoryear{Fritz et al.}{2016}]{Fritz 2016}
Fritz T.~K., Chatzopoulos S., Gerhard O., Gillessen S., Genzel R., Pfuhl O., Tacchella S., et al., 2016, ApJ, 821, 44

\bibitem[\protect\citeauthoryear{Fuller et al.}{2019}]{Fuller 2019a}
Fuller J., Piro A.~L., Jermyn A.~S., 2019, MNRAS, 485, 3661

\bibitem[\protect\citeauthoryear{Fuller \& Ma}{2019}]{Fuller 2019b}
Fuller J., Ma L., 2019, ApJL, 881, L1

\bibitem[\protect\citeauthoryear{Hamers \& Lai}{2017}]{Hamers and Lai 2017}
Hamers A.~S., Lai D., 2017, MNRAS, 470, 1657

\bibitem[\protect\citeauthoryear{Hamers et al.}{2018}]{Hamers 2018}
Hamers A.~S., Bar-Or B., Petrovich C., Antonini F., 2018, ApJ, 865, 2

\bibitem[\protect\citeauthoryear{Hamers \& Safarzadeh}{2020}]{Hamers 2020}
Hamers A.~S., Safarzadeh M., 2020, arXiv, arXiv:2005.03045

\bibitem[\protect\citeauthoryear{Hoang et al.}{2018}]{Hoang 2017}
Hoang B.-M., Naoz S., Kocsis B., Rasio F.~A., Dosopoulou F., 2018, ApJ, 856, 140

\bibitem[\protect\citeauthoryear{Hobbs et al.}{2005}]{Hobbs 2005}
Hobbs G., Lorimer D.~R., Lyne A.~G., Kramer M., 2005, MNRAS, 360, 974

\bibitem[\protect\citeauthoryear{Kocsis \& Tremaine}{2011}]{Kocsis 2011}
Kocsis B., Tremaine S., 2011, MNRAS, 412, 187

\bibitem[\protect\citeauthoryear{Kozai}{1962}]{Kozai} Kozai Y., 1962, AJ, 67, 591

\bibitem[\protect\citeauthoryear{Kiseleva et al.}{1996}]{Stability condition}
Kiseleva L.~G., Aarseth S.~J., Eggleton P.~P., de La Fuente Marcos R., 1996, ASPC, 90, 433

\bibitem[\protect\citeauthoryear{Lidov}{1962}]{Lidov} Lidov M. L., 1962, Planet. Space Sci., 9, 719

\bibitem[\protect\citeauthoryear{Lipunov et al.}{1997}]{Lipunov 1997}
Lipunov V.~M., Postnov K.~A., Prokhorov M.~E., 1997, AstL, 23, 492

\bibitem[\protect\citeauthoryear{Lipunov et al.}{2017}]{Lipunov 2017}
Lipunov V.~M., et al., 2017, MNRAS, 465, 3656

\bibitem[\protect\citeauthoryear{Liu et al.}{2015}]{Liu et al 2015}
Liu B., Mu{\~n}oz D.~J., Lai D., 2015, MNRAS, 447, 747

\bibitem[\protect\citeauthoryear{Liu \& Lai}{2018}]{Liu APJ}
Liu B., Lai D., 2018, ApJ, 863, 68

\bibitem[\protect\citeauthoryear{Liu \& Lai}{2019}]{Liu Quad}
Liu B., Lai D., 2019, MNRAS, 483, 4060

\bibitem[\protect\citeauthoryear{Liu et al.}{2019a}]{Liu APJ 2}
Liu B., Lai D., Wang Y.-H., 2019, ApJ, 881, 41

\bibitem[\protect\citeauthoryear{Liu et al.}{2019b}]{Liu SMBH}
Liu B., Lai D., Wang Y.-H., 2019, ApJL, 883, L7

\bibitem[\protect\citeauthoryear{Liu \& Lai}{2020}]{Liu SMBH 2}
Liu B., Lai D., 2020, PhRvD, 102, 023020

\bibitem[\protect\citeauthoryear{Lousto et al.}{2010}]{Lousto 2010}
Lousto C.~O., Campanelli M., Zlochower Y., Nakano H., 2010, CQGra, 27, 114006

\bibitem[\protect\citeauthoryear{Lu et al.}{2021}]{Lu 2021}
Lu W., Beniamini P., Bonnerot C., 2021, MNRAS, 500, 1817

\bibitem[\protect\citeauthoryear{Mandel}{2016}]{Mandel 2016}
Mandel I., 2016, MNRAS, 456, 578

\bibitem[\protect\citeauthoryear{Mandel \& de Mink}{2016}]{Mandel and de Mink 2016}
Mandel I., de Mink S.~E., 2016, MNRAS, 458, 2634

\bibitem[\protect\citeauthoryear{Mandel \& Fragos}{2020}]{Mandel 2020}
Mandel I., Fragos T., 2020, ApJL, 895, L28

\bibitem[\protect\citeauthoryear{Marchant et al.}{2016}]{Marchant 2016}
Marchant P., Langer N., Podsiadlowski P., Tauris T.~M., Moriya T.~J., 2016, A\&A, 588, A50

\bibitem[\protect\citeauthoryear{Miller \& Hamilton}{2002}]{Miller 2002}
Miller M.~C., Hamilton D.~P., 2002, ApJ, 576, 894

\bibitem[\protect\citeauthoryear{Miller \& Lauburg}{2009}]{Miller 2009}
Miller M.~C., Lauburg V.~M., 2009, ApJ, 692, 917

\bibitem[\protect\citeauthoryear{Moe \& Di Stefano}{2017}]{Moe 2017}
Moe M., Di Stefano R., 2017, ApJS, 230, 15

\bibitem[\protect\citeauthoryear{Mu{\~n}oz et al.}{2016}]{Diego 2016}
Mu{\~n}oz D.~J., Lai D., Liu B., 2016, MNRAS, 460, 1086

\bibitem[\protect\citeauthoryear{Naoz}{2016}]{Naoz 2016}
Naoz S., 2016, ARA\&A, 54, 441

\bibitem[\protect\citeauthoryear{O'Leary et al.}{2006}]{O'Leary 2006}
O'Leary R.~M., Rasio F.~A., Fregeau J.~M., Ivanova N., O'Shaughnessy R., 2006, ApJ, 637, 937

\bibitem[\protect\citeauthoryear{Olejak et al.}{2020}]{Olejak 2020}
Olejak A., Fishbach M., Belczynski K., Holz D.~E., Lasota J.-P., Miller M.~C., Bulik T., 2020, arXiv, arXiv:2004.11866

\bibitem[\protect\citeauthoryear{{\"O}zel et al.}{2010}]{Feryal 2010}
{\"O}zel F., Psaltis D., Narayan R., McClintock J.~E., 2010, ApJ, 725, 1918

\bibitem[\protect\citeauthoryear{Petrovich \& Antonini}{2017}]{Petrovich 2017}
Petrovich C., Antonini F., 2017, ApJ, 846, 146

\bibitem[\protect\citeauthoryear{Pijloo et al.}{2012}]{Pijloo 2012}
Pijloo J.~T., Caputo D.~P., Portegies Zwart S.~F., 2012, MNRAS, 424, 2914

\bibitem[\protect\citeauthoryear{Podsiadlowski et al.}{2003}]{Podsiadlowski 2003}
Podsiadlowski P., Rappaport S., Han Z., 2003, MNRAS, 341, 385

\bibitem[\protect\citeauthoryear{Portegies Zwart \& McMillan}{2000}]{Portegies 2000}
Portegies Zwart S.~F., McMillan S.~L.~W., 2000, ApJ, 528, L17

\bibitem[\protect\citeauthoryear{Randall \& Xianyu}{2018}]{Xianyu 2018}
Randall L., Xianyu Z.-Z., 2018, ApJ, 853, 93

\bibitem[\protect\citeauthoryear{Rauch \& Tremaine}{1996}]{Rauch 1996}
Rauch K.~P., Tremaine S., 1996, NewA, 1, 149

\bibitem[\protect\citeauthoryear{Repetto \& Nelemans}{2015}]{Repetto 2015}
Repetto S., Nelemans G., 2015, MNRAS, 453, 3341

\bibitem[\protect\citeauthoryear{Rodriguez et al.}{2015}]{Rodriguez 2015}
Rodriguez C.~L., Morscher M., Pattabiraman B., Chatterjee S., Haster C.-J., Rasio F.~A., 2015, PhRvL, 115, 051101

\bibitem[\protect\citeauthoryear{Rodriguez et al.}{2020}]{Rodriguez 2020}
Rodriguez C.~L., Kremer K., Grudi{\'c} M.~Y., Hafen Z., Chatterjee S., Fragione G., Lamberts A., et al., 2020, ApJL, 896, L10

\bibitem[\protect\citeauthoryear{Samsing et al.}{2014}]{Samsing 2014}
Samsing J., MacLeod M., Ramirez-Ruiz E., 2014, ApJ, 784, 71

\bibitem[\protect\citeauthoryear{Samsing \& D'Orazio}{2018}]{Samsing 2018}
Samsing J., D'Orazio D.~J., 2018, MNRAS, 481, 5445

\bibitem[\protect\citeauthoryear{Samsing \& Hotokezaka}{2020}]{Samsing 2020}
Samsing J., Hotokezaka K., 2020, arXiv, arXiv:2006.09744

\bibitem[\protect\citeauthoryear{Sana et al.}{2012}]{Sana 2012}
Sana H., de Mink S.~E., de Koter A., Langer N., Evans C.~J., Gieles M., Gosset E., et al., 2012, Sci, 337, 444

\bibitem[\protect\citeauthoryear{Silsbee \& Tremaine}{2017}]{Silsbee and Tremaine 2017}
Silsbee K., Tremaine S., 2017, ApJ, 836, 39

\bibitem[\protect\citeauthoryear{Thompson}{2011}]{Thompson 2011}
Thompson T.~A., 2011, ApJ, 741, 82

\bibitem[\protect\citeauthoryear{Tian et al.}{2020}]{Tian 2020}
Tian H.-J., El-Badry K., Rix H.-W., Gould A., 2020, ApJS, 246, 4

\bibitem[\protect\citeauthoryear{Venumadhav et al.}{2020}]{Venumadhav 2020}
Venumadhav T., Zackay B., Roulet J., Dai L., Zaldarriaga M., 2020, PhRvD, 101, 083030

\bibitem[\protect\citeauthoryear{Wen}{2003}]{Wen 2003}
Wen L., 2003, ApJ, 598, 419

\bibitem[\protect\citeauthoryear{Woosley}{2017}]{Woosley 2017}
Woosley S.~E., 2017, ApJ, 836, 244

\bibitem[\protect\citeauthoryear{Yang et al.}{2020}]{Yang 2020}
Yang Y., Gayathri V., Bartos I., Haiman Z., Safarzadeh M., Tagawa H., 2020, arXiv, arXiv:2007.04781

\bibitem[\protect\citeauthoryear{Zackay et al.}{2019}]{Zackay 2019}
Zackay B., Venumadhav T., Dai L., Roulet J., Zaldarriaga M., 2019, PhRvD, 100, 023007

\bibitem[\protect\citeauthoryear{Zevin et al.}{2020}]{Zevin 2020}
Zevin M., Spera M., Berry C.~P.~L., Kalogera V., 2020, arXiv, arXiv:2006.14573

\bibitem[\protect\citeauthoryear{Ziosi et al.}{2014}]{Ziosi 2014}
Ziosi B.~M., Mapelli M., Branchesi M., Tormen G., 2014, MNRAS, 441, 3703

\end{thebibliography}
\end{document}